\documentclass{aastex}
\usepackage{wasysym}
\usepackage{lscape}
\newcommand{\etal}{et al.}

\begin{document}

\title{Terrestrial Planet Formation Surrounding Close Binary Stars}

\author{Elisa V. Quintana and Jack J. Lissauer}

\affil{Space Science and Astrobiology Division 245-3, NASA Ames Research Center, Moffett Field, CA 94035}

\email {equintan@pollack.arc.nasa.gov}

\begin{abstract} 
Most stars reside in binary/multiple star systems; however, previous
models of planet formation have studied growth of bodies orbiting an
isolated single star.  Disk material has been observed around both
components of some young close binary star systems.  Additionally, it
has been shown that if planets form at the right places within such
disks, they can remain dynamically stable for very long times.

Herein, we numerically simulate the late stages of terrestrial planet
growth in circumbinary disks around `close' binary star systems with
stellar separations 0.05 AU $\leq a_B \leq$ 0.4 AU and binary
eccentricities 0 $\leq e_B \leq$ 0.8.  In each simulation, the sum of
the masses of the two stars is 1 M$_{\odot}$, and giant planets are
included.  The initial disk of planetary embryos is the same as that
used for simulating the late stages of terrestrial planet formation
within our Solar System by Chambers (2001, Making more terrestrial
planets, Icarus 152, 205-224), and around each individual component of
the $\alpha$ Centauri AB binary star system by Quintana et al. (2002,
Terrestrial planet formation in the $\alpha$ Centauri system,
Astrophys. J. 576, 982-996).  Multiple simulations are performed for
each binary star system under study, and our results are statistically
compared to a set of planet formation simulations in the
Sun-Jupiter-Saturn system that begin with essentially the same initial
disk of protoplanets.  The planetary systems formed around binaries
with apastron distances $Q_B$ $\equiv$ $a_B$(1 + $e_B$) $\lesssim$ 0.2
AU are very similar to those around single stars, whereas those with
larger maximum separations tend to be sparcer, with fewer planets,
especially interior to 1 AU.  We also provide formulae that can be
used to scale results of planetary accretion simulations to various
systems with different total stellar mass, disk sizes, and
planetesimal masses and densities.
\end{abstract}

\textit{Keywords:} Planetary formation; Terrestrial planets; Extrasolar planets

\section{Introduction} 
More than half of all main sequence stars, and an even larger fraction
of pre-main sequence stars, are in binary/multiple star systems
(Duquennoy and Mayor 1991; Mathieu \etal\ 2000).  Virtually all
previous models of planet formation, however, have assumed an isolated
single star.  Of the first 131 extrasolar planet systems that have been
confirmed, at least 30 are on so-called S-type orbits that encircle one
component of a binary star system, including at least 3 that orbit one
member of a triple-star system (Raghavan \etal\ 2006).  The effect of the stellar companion on the formation of these planets,
however, remains unclear.

One planet has been detected in a P-type orbit which encircles both
members of a binary star system.  This planet, which has a minimum
mass of $\sim$ 2.5 times the mass of Jupiter (M$_{\jupiter}$), orbits
$\sim$ 23 AU from the center of mass of PSR 1620-26, a radio pulsar
binary comprised of a neutron star and a white dwarf in a $\sim$ 191
day stellar orbit (Lyne \etal\ 1988, Sigurdsson 1993, Sigurdsson
\etal\ 2003).  The most plausible model for its formation is accretion
within a metal-rich disk produced by post-main sequence Roche lobe
overflow (Lissauer 2004).  Planets have not been detected in a P-type
orbit around two main sequence stars, but short-period binaries are
not included in precise Doppler radial velocity search programs
because of their complex and varying spectra.  Planets in P-type
orbits around the eclipsing binary star system CM Draconis have been
searched for using the eclipse timing variation method (Deeg
\etal\ 2000), but results were not definitive.  Two substellar
companions have been detected around the G6V star HD 202206, with
minimum masses of 17.4 M$_{\jupiter}$ (at 0.83 AU) and 2.44
M$_{\jupiter}$ (at 2.55 AU) (Udry \etal\ 2002).  The inner companion
is so massive that it is considered to be a brown dwarf, and it is
likely that the outer companion formed from within a circumbinary
(star-brown dwarf) disk (Correia \etal\ 2005).  A more general
discussion of the detectability of circumbinary planets is presented
by Muterspaugh (2005).  Note also that the observation of two small
moons orbiting in nearly circular/planar orbits about Pluto-Charon
(Weaver \etal\ 2006), a system which is like a binary with an 8/1 mass
ratio, suggests that accretion can occur in P-type orbits about close
binaries.

The main objective of this article is to numerically examine the late
stages of terrestrial planet formation around both members of a binary
star system.  The existence of Earth-like planets in orbit about one
or both components of main sequence binary stars has yet to be
determined, though ground- and space-based efforts to search for
extrasolar terrestrial planets are currently in development.  An
additional benefit of understanding the differences between planet
formation around single stars and that around close binaries is that
for eclipsing binaries, the contrast ratio between brightness of the
stars and that of the planet(s) is reduced during the eclipse.  For a
total eclipse of identical stars, this reduction is a factor of two;
as lower mass main sequence stars can be just slightly smaller but
significantly less luminous, the detectability of the planet can be
enhanced by more than a factor of two when the fainter star transits
the brighter one.  In an evolved close binary having undergone mass
transfer, the fainter star can actually completely eclipse its much
brighter companion, leading to an even larger improvement in planetary
detectability.

In the conventional model of planet formation, terrestrial planets are
believed to have formed by an accretion process from within a disk of
gas and dust that has remained around a newly formed star (Safronov
1969, Lissauer 1993).  The coexistence of disks of material with stars
that possess a stellar companion support the idea that planet
formation within binary star systems may be common. Circumbinary disk
material has been detected through millimeter and mid-infrared excess
emission around several spectroscopic pre-main sequence binary star
systems with stellar semimajor axes $a_B$ $\lesssim$ 1 AU.  These
systems include GW Ori (Mathieu \etal\ 1995), UZ Tau E (Jensen
\etal\ 1996), and DQ Tau (Mathieu \etal\ 1997).  The masses of these
disks are each comparable to or exceed the minimum mass of the solar
nebula, $\sim$ 0.01 solar mass (M$_{\odot}$) (Weidenshilling 1977),
and are also comparable to the masses of disks found around single
stars.

Numerical models of circumbinary disks find that, for binary star
systems with binary eccentricities ($e_B$) increasing from 0 -- 0.25,
the inner edge of a gaseous disk is truncated to within $\sim$ 1.8 --
2.6 times the semimajor axis of the binary stars' mutual orbit ($a_B$)
(Artymowicz and Lubow 1994, Lubow and Artymowicz 2000).  A star with a
giant planet orbiting interior to the terrestrial planet region is
dynamically a binary system of extreme mass ratio.  Raymond
\etal\ (2005) showed that planetary embryos can accrete into
terrestrial planets around a star that has a close-in (between 0.15 AU
-- 0.5 AU) Jupiter-mass planet.

Herein, we simulate terrestrial planetary accretion within a
circumbinary disk of protoplanets around `close' ($a_B$ = 0.05 -- 0.4
AU) binary star systems that each have a combined stellar mass of 1
M$_{\odot}$.  Our numerical method and the initial states of the
systems that we have simulated are given in Section 2.  Section 3
examines the regions of stability for test particles orbiting about
these binary star systems.  The results of the close binary accretion
simulations, including a quantitative analysis of the final planetary
systems formed, are presented in Section 4.  Our conclusions are
discussed in Section 5.  Appendix A presents new simulations of the
late stages of terrestrial planet formation in the Sun-Jupiter-Saturn
system that we have performed to facilitate comparisons between planet
growth around single and close binary stars, and simulations using an
initial disk of bodies whose eccentricities are forced by the binary
stars are presented in Appendix B.  In Appendix C, we discuss the
scaling of our results to systems with different planetesimal
densities, disk sizes, and stellar masses.

\section{Initial Conditions and Numerical Model} 
The combined mass of the binary stars is equal to 1 M$_\odot$ in all
of the simulations, with the stellar mass ratio $\mu$ (the ratio of
the secondary star's mass to the total stellar mass) equal to either
0.2 or 0.5.  Binary star separations in the range $a_B$ = 0.05 AU --
0.4 AU are examined, while $e_B$ begins at 0, 1/3, 0.5, or 0.8 such
that the stellar apastron $Q_B \equiv a_B(1 + e_B)$ is 0.05 AU $\leq
Q_B \leq$ 0.4 AU.  Not all combinations of these stellar parameters,
however, are used.  For most of the simulations, the midplane of the
circumbinary disk begins coplanar to the stellar orbit, but for one
set of binary star parameters a relative inclination of $i$ =
30$^{\circ}$ is investigated.  Although a stellar companion present
during the earlier stages of planet formation would likely force the
planetesimal disk into the plane of the binary orbit, many binary
stars may originate as unstable triple star systems which could
produce a binary star system with an accretion disk at a high relative
inclination.  It is also possible that a companion may have been
captured around a single star that posesses an accretion disk.

Giant planets are included in the simulations, as they are in most
simulations of the late stages of terrestrial planet accumulation in
our Solar System (Chambers 2001, Appendix A).  In all of the planetary
accretion simulations presented herein, a Jupiter-mass planet is
placed in the system at $a_{\jupiter}$ = 5.2 AU, with an eccentricity
of $e_{\jupiter}$ = 0.048 and an inclination of $i_{\jupiter}$ =
0.36$^{\circ}$ relative to the midplane of the disk.  Apart from a
single set of runs in which the stars are separated by $a_B$ = 0.05 AU
and travel on an initially circular orbit coplanar with the midplane
of the disk, a second giant planet of Saturn's mass, with
$a_{\saturn}$ = 9.54 AU, $e_{\saturn}$ = 0.053, and $i_{\saturn}$ =
0.89$^{\circ}$ relative to the midplane of the disk, is also included.
The effect of the stellar pertubations on these giant planets is
discussed in Section 4.2.

\subsection{Circumbinary Disk Model} 
The initial conditions for the bodies in the circumbinary disk are
based upon earlier numerical simulations of the late stages of
terrestrial planet formation in the Sun-Jupiter-Saturn (SJS) system
(Chambers 2001) which resulted in the formation of planetary systems
with masses and orbits similar to the terrestrial planets in the Solar
System (see Appendix A).  In this model, 14 planetary embryos (each
with a mass of 0.0933 times the mass of the Earth, M$_{\oplus}$) are
embedded in a disk of 140 smaller planetesimals (each with a mass of 0.00933 M$_{\oplus}$), all lying between 0.36 AU and 2.05 AU of the center of mass of
the binary stars.  The radii of these bodies are calculated assuming a
material density of 3 g cm$^{-3}$.  All other initial orbital elements
were chosen at random, with the eccentricities ranging from 0 -- 0.01,
and inclinations relative to the mean plane of the disk $\leq$
0.5$^{\circ}$.  Analyses of trajectories around close binaries show that closed, near
circular, non-crossing orbits occur in situations such as this
(Pichardo \etal\ 2005, see especially their Figures 6 and 9). In Appendix B, we present one set of simulations that was run
using a disk of planetesimals and embryos with initial eccentricities
determined by perturbations from the binary.

In the majority of our simulations, the initial planetesimal/embryo
disk extends closer to the stars than the region in which
planetesimals are expected to be able to form within a {\it gas-free}
disk (Moriwaki and Nakagawa 2004).  However, calculations show that
planetesimal growth occurs over a much greater region within {\it
  gas-rich} circumstellar (S-type orbits) disks than within analogous
regions of gas-free disks (Kortenkamp and Wetherill 2000, Thebault \etal\ 2006), and we
would expect an analogous situation for P-type orbital regions.
Moreover, it is also possible that at least the initial phases of
planetesimal growth can occur farther from the stars, and the
planetesimals can then migrate inwards as a consequence of gas drag.

\subsection{Numerical Method} 
To examine both the orbital stability and accretion of bodies in a
disk, we use one of the two symplectic $N$-body algorithms that we
developed to examine planetary accretion in binary star
systems (Chambers \etal\ 2002).  The `close binary' algorithm (used in
the simulations presented in this article) was designed to examine
accretion in P-type orbits around binary stars.  The `wide binary'
algorithm, which follows bodies in S-type orbits within binary star
systems, was recently used to simulate the late stages of terrestrial
planet formation around each star of the nearest binary star system to
the Sun, $\alpha$ Centauri AB (Quintana et al. 2002, Quintana 2003).
In these simulations, which began with an initial disk mass
distribution virtually identical to the disk discussed above, from 3
-- 5 terrestrial planets formed on stable orbits around each
individual component in $\alpha$ Cen AB provided the initial
inclination of the disk relative to the stellar orbit began at or
below $\sim$ 30$^{\circ}$.  Numerous simulations of terrestrial planet
formation in S-type orbits in main sequence binary star systems, with
the aim of examining a larger binary star parameter space, will be
presented in Quintana \etal\ (2007).

The close binary algorithm calculates the temporal evolution of the
position and velocity of each body in the disk with respect to the
center of mass of the binary stars, subject to gravitational
perturbations from both stars and to gravitational interactions and
completely inelastic collisions among the bodies.  Bodies are removed
if their orbit extends more than 100 AU from the more massive star, or
if they orbit too close to the center of mass of the binary stars.
For selected simulations with larger stellar separations, material is
removed if its distance from the center of mass exceeds the smaller star's apastron distance by less than 0.1 AU.  A time-step of 7
days is used for the bodies in the disk, while the binary stars are
given a time-step that is shorter by approximately the ratio of the
binary period to the orbital period of the innermost planetesimal (see
Chambers et al. 2002 for details).  The period of the innermost body
is 55.09 days, so the choice of a 7 day time-step may not accurately
follow the evolution of the innermost bodies, which could lead to
errors, especially concerning the amount of mass lost inside.  We are
comparing our results, however, to the simulations in the
Sun-Jupiter-Saturn system (Chambers 2001) and to simulations around
each star in $\alpha$ Cen AB, each of which used a 7-day time-step, so
the differences that we find should be real.\footnote{To investigate the statistical validity of simulations
  performed using a 7 day time-step, we performed test particle
  simulations for five of the binary star systems that are examined in
  this article using time-steps of 3.5 days, 3.6 days, and 7 days for
  each system.  For the binary star systems with with stellar
  parameters ($a_B$, $e_B$, $\mu$) equal to (0.05, 0, 0.5) and (0.1,
  0.8, 0.5), from 25 -- 50 test particles were placed around each
  system at 0.01 AU intervals, beginning at the distance from the
  center of mass of the stars at which bodies are removed.  For the
  binary star systems with ($a_B$, $e_B$, $\mu$) equal to (0.2, 0.5,
  0.5), (0.3, 1/3, 0.5), and (0.4, 0, 0.5), the smallest semimajor
  axis for which bodies can be stable was determined, $a_c$ (described
  further in Section 3), and test particles were placed between ($a_c$
  -- 0.25 AU) and ($a_c$ + 0.25 AU), with 0.01 AU intervals.  All
  other orbital elements were chosen at random, with eccentricities $e
  \leq$ 0.01 and inclinations $i \leq$ 0.01 radian, but were kept the
  same for each set of simulations.  The giant planets were not
  included and the test particle orbits were followed for 10 Myr.  For
  the binary system with $a_B$ = 0.05 AU that began on an initially
  circular orbit, the orbital elements of the surviving test particles
  were nearly identical.  Particles in other systems, especially close
  to the stability limit, showed larger variations.  The differences
  in results between the runs with 3.5 day time-steps and 3.6 day
  time-steps, however, were almost as large as the differences from
  the 7 day time-step runs, indicating that the dominant source of
  variation was chaos rather than systematic inaccuracy of the
  integrations.  Moreover, in many of our simulations the innermost
  planetesimal survived the entire integration intact, with little
  variation in its principal orbital elements (as was the case for
  some of our simulations of planetary growth; see, for example,
  Fig. 1).  We therefore conclude that the use of a 7 day time-step
  does not significantly degrade the statistical validity of our
  integrations. }

Because these $N$-body systems are chaotic, each binary star system
under study is simulated five or six times with slightly different
initial conditions for the circumbinary disk.  Of the 154 rocky
embryos and smaller planetesimals, one planetesimal near 1 AU is
initially displaced by 1, 2, 3, 4, or 5 meters along its orbit with
all other parameters within a given set remaining the same.  The
simulations are labelled as follows: CB\_$a_B$\_$e_B$\_$\mu$\_$x$,
where $a_B$ is the binary semimajor axis, $e_B$ is the binary
eccentricity, $\mu$ is the stellar mass ratio, and $x$ (= a, b, c, d,
e, or f) signifies each realization of a given system.  If the
midplane of the initial planetesimal disk is inclined relative to that
of the binary star orbital plane, the inclination angle is listed
between $\mu$ and $x$, but when they are coplanar (as for most of our
simulations), no value is given.  The first run in each set (the ``a''
integrations) begins with the `standard' bi-modal circumbinary disk
mass distribution.  In the remaining runs within a set, a planetesimal
near 1 AU is initially displaced by 1 meter along its orbit (for the
``b'' integrations), by 2 meters along its orbit (for the ``c''
integrations), etc.  The evolution of the material in the disk is
initially followed for 200 Myr.  If it appeared that further
collisions among the planets were fairly likely, individual
integrations were continued for total simulation times of 500 Myr (or
550 Myr in one case) or 1 Gyr.  In simulations that resulted in the
formation of just one planet, the integrations were stopped after the
last collision or ejection.  The results of each accretion simulation
are presented in Table 2 and discussed in Section 4.

\section{Orbital Stability Around Close Binary Star Systems}
The dynamical stability of test particles in P-type orbits has been
previously examined for binary star systems with $\mu$ ranging
from 0.1 -- 0.5 and binary eccentricities $e_B$ between 0.0 -- 0.7
(Holman and Wiegert 1999).  In that study, test particles were placed
in the binary orbital plane between 1.0 $a_B$ and 5.0 $a_B$ (with 0.1
$a_B$ increments) at eight equally spaced longitudes per semimajor
axis, and the system was evolved forward in time for 10$^4$ binary
periods.  The closest distance to the binary at which all eight
particles survived (the `critical semimajor axis', $a_c$) was
calculated for each system.  Depending on $e_B$ and $\mu$, $a_c$ was
found to lie within 2.0 $a_B$ -- 4.1 $a_B$.  Many of these simulations
also revealed an unstable region beyond $a_c$ that corresponded to one
of the system's $n$:1 mean motion resonances, followed by an
additional outer region of stability.

We performed a similar analysis for the close binary
configurations examined herein, with the integration time extending to
10$^6$ binary orbital periods.  Test particles were placed between 1.8
$a_B$ and 5.0 $a_B$ (with 0.1 $a_B$ increments) at eight equally
spaced longitudes.  Integrations were performed using the close binary
algorithm.  Particles were removed from the simulation if they fell
within 0.01 $a_B$ from the more massive star or if their distance
exceeded 100 $a_B$.  For each system, Table 1 gives the smallest
semimajor axis at which bodies at all 8 longitudes survive the
integration, $a_{c}$.  In the close binary systems with $e_B$ = 1/3,
regions of instability were found beyond $a_{c}$, and the minimum semimajor axis at and beyond which all test particles survive is listed in
brackets in Table 1.  Analogous integrations were performed with
finer increments of 0.01 $a_B$ to find the innermost semimajor axis
for which at least one body survived, given by $a_{c}^\ast$ in Table
1.  These results are consistent with those of Holman and Wiegert
(1999), who found a (roughly) linear dependence of the critial
semimajor axis on $e_B$.  Note that for a given $e_B \ge$ 1/3, the
region cleared of test particles is greater for $\mu$ = 0.2 than it is
for $\mu$ = 0.5, presumably because the apastron distance of the
smaller star from the center of mass of the system is larger for $\mu$
= 0.2.  For each close binary accretion simulation, the ratio of the
semimajor axis of the innermost final planet ($a_p$) to the innermost stable
orbit of the system ($a_{c}^\ast$) is presented in Table 2.

\section{Planetary Accretion Simulations}
Figure 1 displays the accretion evolution of system
CB\_.05\_0\_.5$^\dag$\_a (the dagger signifies that only one gas
giant, a Jupiter-mass planet at 5.2 AU, is included). The eccentricity
of each body is shown as a function of semimajor axis, and the radius
of each symbol is proportional to the radius of the body that it
represents.  Throughout the simulation, the larger embryos remain on
orbits with $e$ $\lesssim$ 0.1, whereas the planetesimals become more
dynamically excited with time during the first $\sim$ 10 -- 20 Myr.
Between $\sim$ 5 -- 50 Myr, a trend occurs in which planetesimal
eccentricities increase with increasing semimajor axis as a
consequence of perturbations by Jupiter.  All but one of the
planetesimals are either swept up by the larger embryos or are ejected
from the system.  The first planetesimal ejection occurs at $\sim$ 12
Myr, while the only lost embryo was ejected at $\sim$ 37 Myr.  No
bodies in this system traveled closer to the binary orbit than the
initial semimajor axis of the innermost planetesimal; indeed, the
innermost planetesimal was the one that survived without impact.
After 500 Myr elapsed, 6 planets with masses between 0.11 -- 0.61
M$_{\oplus}$ orbited within 2.3 AU; these planets incorporated 84\% of
the initial disk mass.  Despite the apparent crowding, the system
appears to be quite stable; no bodies are lost or accreted between
92.7 Myr and the end of the simulation.

Figure 2 (CB\_.05\_0\_.5$^\dag$\_d) shows the growth of planets formed
around a binary star system identical to that shown in Fig. 1, but in
this case the initial disk mass is slightly different (one
planetesimal near 1 AU is shifted by 3 meters along its orbit).  Note
that the evolution of the disk in this figure (and subsequent figures)
is shown beginning at $t$ = 0.2 Myr, as the plotted properties of the
disk at $t$ = 0 are identical to those shown in the first panel in
Fig. 1.  The stellar and giant planet perturbations have a similar
effect on the disk as in the simulation shown in Fig. 1.  In this
case, three terrestrial-mass planets have formed within 1.5 AU, while
one planetesimal remains at 2.2 AU, all together incorporating 81\% of the
initial mass in the disk.  The differences in the final planets formed
from two simulations with almost identical initial conditions (Fig. 1
vs. Fig. 2) demonstrate the chaotic nature of these $N$-body systems.

Figure 3 shows run CB\_.05\_0\_.5\_c, which included a Saturn-like
giant planet in addition to the Jupiter-like one.  Nonetheless, the
outcome looks intermediate between the systems shown in Fig. 1 and
Fig. 2.  A comparison between all of the CB\_.05\_0\_.5$^\dag$ runs
with the CB\_.05\_0\_.5 runs (Table 2 and Fig. 7) suggests that the
extra perturbations of ``Saturn'' may reduce the expected number of
terrestrial planets formed, but the effects are small enough that
there is considerable overlap among the chaos-broadened distributions
of outcomes.  Note that these distributions also overlap the results
of simulations of terrestrial planet growth around a single star with
one or two giant planets (Chambers 2001, Chambers \etal\ 2002,
Appendix A) and around the individual stars in the $\alpha$ Centauri
AB binary, provided the disk begins close to the $\alpha$ Cen binary
orbital plane (Quintana \etal\ 2002, Quintana 2003).

Figure 4 displays the evolution of CB\_.075\_.33\_.5\_b, in which the
stars orbit one another on more distant and eccentric paths.  The
system evolves very differently from any of the systems with $a_B$ =
0.05 and $e_B$ = 0.  The innermost large planet is quite eccentric by
$t$ = 20 Myr, and it accretes or scatters all of the smaller bodies
inwards of 1 AU prior to being ejected itself at 106 Myr.  Run
CB\_.075\_.33\_.5\_d produces a similar planetary system, whereas runs
CB\_.075\_.33\_.5\_a, CB\_.075\_.33\_.5\_c, and CB\_.075\_.33\_.5\_e
result in planetary systems resembling those formed in the $a_B$ =
0.05 and $e_B$ = 0 simulations (Table 2 and Fig. 8).  In a sense, this
change in stellar parameters yields systematic differences comparable
to the scatter resulting from chaos.

Figures 5 and 6 show the evolution of systems CB\_.1\_.8\_.2\_d and
CB\_.4\_0\_.5\_c, respectively.  In each case, the binary apastron
distance is much larger than in the runs discussed above, and binary
perturbations clear the system of all but one planet.  Additional
planets remain in the runs with nearly identical initial parameters,
but the systems (Table 2, Figs. 7 and 8) still look much sparcer,
especially in the inner regions, than those formed around a single
star, very close binaries, or individual stars in the $\alpha$ Cen AB
system.  In these cases, systematic effects resulting from the
different binary parameters exceed typical chaotic variations.
Figures showing the temporal evolution of most of the simulations
discussed in this paper are presented in Quintana (2004); plots of
simulations CB\_.1\_0\_.5\_c and CB\_.2\_.5\_.5\_a are presented in
Lissauer \etal\ (2004).

Figures 7 and 8 show the final planetary systems formed in all of our
simulations.  The top left row in each figure shows the Solar System's
terrestrial planets (labelled `MVEM'), followed by the 5 -- 6
realizations of each binary system under study, presented in order of
increasing $Q_B$. Figure 7 presents the final planetary systems formed
around binary stars that began on circular orbits with the disk
initially coplanar to the stellar orbit (labelled in the following
format: $a_B$\_$e_B$\_$\mu$), whereas Fig. 8 shows the results for
simulations with $i$ = 30$^{\circ}$ and the sets of runs with $e_B >
0$.  In these figures, the radius of each body is proportional to the
radius of the planet that it represents, the horizontal lines through
each body indicate the periastron and apastron distances to the center
of mass of the binaries (or the Sun in the MVEM case), the vertical
lines represent the inclination relative to the binary orbital plane
(the Laplacian plane in the MVEM case), and the arrows show the
orientation of the final spin axes of each planet (arrows are omitted
for planetesimals and embryos that survived the integration without a
collision).  Although the final planetary systems formed vary widely
among a given set of binary star parameters due to the chaotic nature
of these simulations, general trends are apparent in the planets
formed around stars with larger separations and higher eccentricities.
In order to quantitatively analyze these effects, we developed a set
of formulae that characterize the orbits and distribution of mass for
all of the final planetary systems.  These are described in the next
subsection and the statistical variations are discussed in subsection
4.2.

\subsection{Parameters and Statistics}
The results of all of our close binary simulations are given in Table
2, which lists the stellar parameters/initial conditions and gives the
values of statistical parameters that were developed to help
characterize the final planetary systems.  Most of these statistics
were previously used to compare the outcomes from accretion
simulations in the Sun-Jupiter-Saturn (SJS) system (Chambers 2001, Appendix
A) and around each star in the $\alpha$ Cen AB binary star system
(Quintana \etal\ 2002, Quintana 2003), all of which used essentially
the same initial planetesimal disk.  The first column lists the name
of each close binary simulation (CB\_$a_B$\_$e_B$\_$\mu$\_$x$, as
described in Section 2).  Simulations that include just one giant
planet (Jupiter) are labelled with $^\dag$; all others include a
Jupiter-like planet and a Saturn-like planet.  When the initial
midplane of the circumbinary disk is inclined relative to the stellar
orbit, the runs are denoted CB\_$a_B$\_$e_B$\_$\mu$\_$i$\_$x$.  The
duration of each simulation is listed in column 2.  Columns 3 -- 15
present the following statistics (see Chambers 2001 and Quintana
\etal\ 2002 for mathematical descriptions of the statistics given in
Columns 7 -- 15; statistics presented in columns 5 and 6 are new in
this work).

\begin{description}
\item[(3)] The number of planets, $N_p$, that are at least as massive
  as the planet Mercury ($\sim$ 0.06 M$_{\oplus}$).  Note that the 14
  planetary embryos in the initial disk each satisfy this mass
  requirement, as do bodies consisting of at least 7 planetesimals.

\item[(4)] The number of minor planets, $N_m$, that are less massive
  than the planet Mercury.

\item[(5)] The ratio of the semimajor axis of the innermost final
  planet to the closest stable orbit of the system,
  $a_p$/$a^{\ast}_{c}$.  Note that in principle, the value of this
  quantity may be (slightly) less than unity, as $a^{\ast}_{c}$ was
  estimated using a coarse grid and only considered bodies initially
  on circular orbits within the binary plane.

\item[(6)] The ratio of the periastron of the innermost final planet,
  $q_p$ = $a_p$(1 -- $e_p$), to the binary apastron $Q_B$ = $a_B$(1 +
  $e_B$).

\item[(7)] The fraction of (the final) mass in the largest planet, $S_m$.

\item[(8)] An orbital spacing statistic, $S_s$, which gives a measure
  of the distances between the orbits of the final planets (that are
  larger than the planet Mercury).  Larger values of $S_s$ imply more
  widely spaced final planets.

\item[(9)] The normalized angular momentum deficit, $S_d$, which
  measures the fractional difference between the planets' actual
  orbital angular momenta and the angular momenta that they would have
  on circular, uninclined orbits with the same semimajor axes.

\item[(10)] A mass concentration statistic, $S_c$, which measures the
  degree to which mass is concentrated in one part of the planetary
  system.

\item[(11)] A radial mixing statistic, $S_r$, which sums the radial
  migrations of the bodies that form a planet.

\item[(12)] The percentage of the initial mass that was lost from the planetary
  system (came too close to the stars or was ejected to interstellar space), $m_l$.

\item[(13)] The total mechanical (kinetic + potential) energy per unit
  mass for the planets remaining at the end of a simulation, $E$,
  normalized by $E_{0}$, the energy per unit mass of the system prior
  to the integration.

\item[(14)] The angular momentum per unit mass of the final planets,
  $L$, normalized by $L_0$, the angular momentum per unit mass of the
  initial system.

\item[(15)] The $Z$ component of angular momentum per unit mass
  relative to the stellar orbit, ${L}{_Z}$, normalized by ${L_0}_Z$,
  the initial $Z$ component of angular momentum of the system.

\end{description}

Following the close binary results in Table 2 are analogous statistics
for the following systems: the four terrestrial planets in the Solar
System (labelled `MVEM'); the averaged values for 31 accretion
simulations in the Sun-Jupiter-Saturn system (`SJS$\_{\rm{ave}}$',
which are presented in Appendix A); the averaged values of a set of
accretion simulations around the Sun with neither giant planets nor a
stellar companion perturbing the system (`Sun$\_{\rm{ave}}$', Quintana
\etal\ 2002, Appendix A); the averaged values for the planets formed
within 2 AU of the Sun when neither giant planets nor a stellar
companion is included (`Sun$\_{\rm{ave}}$ ($a <$ 2 AU)', Quintana
\etal\ 2002, Appendix A); and the averaged values for the planetary
systems formed around $\alpha$ Cen A in simulations for which the
accretion disk began with an inclination $i \leq$ 30$^{\circ}$ to the
$\alpha$ Cen AB binary orbital plane (labelled `$\alpha$ Cen ($i \leq$
30$^{\circ}$)', Quintana \etal\ 2002).  Note that only the six
statistics listed for the terrestrial planets in our Solar System
(MVEM) are actual observables.

We use a two-sided Kolmogorov-Smirnov (K-S) test to compare each
planetary statistic for each binary star configuration to the
analogous planetary statistic from the distribution of 31 SJS
simulations (Appendix A).  Table 2 gives the K-S statistic, D, and the
associated probability, P, for each set of simulations.  Generally,
values of P $\lesssim$ 0.05 indicate that the two sets of data are
drawn from different distribution functions, $\it{i.e.}$, the effect
of the binary stars on the disk is statistically significant.  The
orbital spacing statistic, $S_s$ (column 8 in Table 2), is undefined
in simulations which resulted in the formation of a single planet
($N_p$ = 1, including those systems with $N_m$ $\neq$ 0 since these
smaller bodies are neglected for this calculation).  Simulations with
$N_p$ = 1 and $N_m$ = 0 have an infinite value for the mass
concentration statistic, $S_c$ (column 7 in Table 2).  In each case,
when calculating D and P, the values for these statistics are replaced
by the highest finite value of the statistic within that set of runs
in order to minimize biasing of the results.  A discussion of the
results from Table 2 are presented in the next subsection.

\subsection{Statistical Variations Among the Systems}
Nearly all of the simulations that began with binary stars with $Q_B
\leq$ 0.2 AU resulted in distributions of planetary systems that are
statistically consistent in most properties with those formed in the
SJS simulations.  The mass loss, final specific energy, and final
specific angular momentum statistics for the close binary simulations
(columns 12 -- 15 in Table 2), however, differ from the corresponding
SJS distributions, and will be discussed later in this section.  One
set of simulations (the CB\_.1\_0\_.5 runs) in which 4 -- 5
terrestrial-mass planets formed (compared to an average of 3 planets
formed in the SJS runs) have planetary statistics that are
inconsistent with the SJS distributions, even though the final planets
have masses and orbits that appear upon inspection to be similar to
the terrestrial planets in our Solar System.  This divergence is
possible because the statistical tests are of marginal use for
comparing ensembles with only five members.  We include the statistics
because they provide a different, albeit not necessarily better,
perspective from a visual comparison of the final systems shown in
Figures 7 and 8.

Neglecting to include a Saturn-like planet in addition to the
Jupiter-like planet in the simulations did not affect the final
outcomes of the planets in a statistically significant manner.  The
first two sets of runs listed in Table 2 ($a_B$ = 0.05) show that the
effects of chaos are larger than the effects from the number of giant
planets that are included.  Nonetheless, it is worth noting that the
most crowded final system, run CB\_.05\_0\_.5$^\dag$\_a, in which 6
planets and 1 planetesimal survived, was subject to the perturbations
of only a single giant planet.

One surprising result is that one run ended with 6 terrestrial planets
and five runs concluded with 5 planets.  In all of these runs, the
binary stars' mutual orbit was circular with $a_B$ = 0.05 AU, 0.1 AU
or 0.2 AU.  In contrast, at most 4 planets remained in the single star
simulations listed in Table 3.  Within each $\it{set}$ of close binary
simulations, at least one planet more massive than the planet Mercury
remained in an orbit farther from the center of mass of the system
than the present orbit of Mars ($\sim$ 1.5 AU).  In many of the close
binary simulations, Mars-sized planets formed (and/or planetesimals
remained) in orbits beyond $\sim$ 2 AU from the center of mass of the
binary stars.  In our Solar System and in the SJS simulations, the
location of the $\nu_6$ secular resonance restricts terrestrial planet
formation to within $\sim$ 2 AU of the Sun.  In contrast, orbital
precession induced by the binary displaces secular resonances of the
giant planets away from the terrestrial planet zone.  We integrated
test particles orbiting 1.5 -- 2.5 AU from a single or close binary
star that had Jupiter and Saturn-like planetary companions.  The
eccentricities excited in the test particles orbiting the single star
reached substantially higher values than did those of test particles
around the binary, consistent with expectations.

The variations in the orbital elements of Jupiter and Saturn are
larger in most of the close binary simulations compared to simulations
with a single star.  In the SJS set, the average peak to trough
variations of Jupiter's semimajor axis, eccentricity, and inclination
are $\Delta a_{\jupiter} \sim$ 0.006 AU, $\Delta e_{\jupiter} \sim$
0.038, and $\Delta i_{\jupiter} \sim$ 0.32$^{\circ}$, respectively.
For Saturn, these variations are $\Delta a_{\saturn} \sim$ 0.077 AU,
$\Delta e_{\saturn} \sim$ 0.086, and $\Delta i_{\saturn} \sim$
0.785$^{\circ}$.  In the close binary simulations, the average peak to
trough variations in semimajor axes and eccentricities for both
Jupiter and Saturn are slightly larger (although the changes are more
chaotic systems with increasing $Q_B$), with $\Delta a_{\jupiter}
\sim$ 0.01 AU, $\Delta e_{\jupiter} \sim$ 0.06, $\Delta a_{\saturn}
\sim$ 0.14 AU, and $\Delta e_{\saturn} \sim$ 0.11.  The variations in
inclination increase with increasing $Q_B$ for both Jupiter (up to
2.3$^{\circ}$) and Saturn (up to 3.5$^{\circ}$).

In the majority of simulations that began with binary stars on
circular orbits and $a_B \leq$ 0.2 AU, the terrestrial planet systems
that formed span essentially the entire range of the initial disk.
The stellar perturbations on the inner edge of the disk become
apparent in simulations with $Q_B \geq$ 0.3 AU and in many of the
simulations with smaller $Q_B$ but $e_B \textgreater$ 0 (Fig. 7 and
Fig. 8).  The statistics in columns 5 and 6 of Table 2 give a measure of
how close the innermost planet forms to the innermost stable orbit of
the system, $a_p$/$a^{\ast}_{c}$, and the ratio of the closest
approach of the planet to the binary apastron, $q_p$/$Q_{B}$.  These
statistics do not have analogs in the SJS system, but are useful in
comparing simulations with differing binary star parameters.  Note
that for most final systems, $q_p$/$Q_{B}$ $\sim$ 2
$a_p$/$a^{\ast}_{c}$.  In the first two sets of simulations ($a_B$ =
0.05 AU), the stars have a minimal effect on the inner edge of the
disk, which begins at more than 3.5 times the distance (from the
center of mass of the binary stars) of the location of the innermost
stable orbit of the system, $a^{\ast}_{c}$.  In more than one-third of
the simulations with $Q_B \leq$ 0.1 AU (all of
which use $e_B \leq$ 1/3 and have $a^{\ast}_{c}$ $\lesssim$ 0.2 AU), the innermost
planetesimal in the initial disk survived the entire integration
without a collision and remained close to its initial orbit at $\sim$ 0.35
AU (Fig. 7 and 8).  In eight of the ten simulations with $a_B$ = 0.2
and $e_B$ = 0, the innermost embryo survived the integration without a
collision.  Binary systems with $Q_B \geq$ 0.18 AU have larger values
of $a^{\ast}_{c}$ (0.32 AU $\textless$ $a^{\ast}_{c}$ $\textless$ 0.9 AU) that
approach or exceed the initial orbit of the innermost body in the
protoplanetary disk.  Although the inner edge of the disk is
truncated, the innermost planets formed in many of these simulations
remain on orbits that are close to the system's $a^{\ast}_{c}$.

In the SJS set of simulations and in the CB\_.05\_0\_.5$^\dag$ runs,
an average of $\sim$ 51\% of the mass that composes the final planets
remained in the largest planet formed, the same fraction of mass that
composes the Earth in the Mercury-Venus-Earth-Mars system.  The value
of $S_m$ (the fraction of final mass that composes the largest planet)
is generally higher in the close binary simulations that resulted in
fewer planets.  With the exception of two sets of runs (the
CB\_.1\_0\_.5 set with an average value of $S_m$ = 0.35, and the
CB\_.15\_1/3\_.5 set with an average of $S_m$ = 0.64), all of the sets
of simulations with $Q_B \leq$ 0.2 AU have values of $S_m$ that are
statistically consistent with those in the SJS set of runs (typically
$\sim$ 0.5).  The value of $S_m$ tends to be larger for planets that
form around more eccentric binary stars, yet the effect of varying the
stellar mass ratio for systems with common $a_B$ and $e_B$ appears to
be statistically insignificant.

The statistic $S_s$, which quantifies the orbital spacing of the final
planets, ranged from 29 -- 87 (with an average of 44) in the SJS
distribution, and is $S_s$ = 38 for the terrestrial planets in the
Solar System.  The CB\_.1\_0\_.5 simulations resulted in a higher
number of planets that were more closely spaced, with $S_s$ = 32 on
average.  All other sets of close binary simulations with $Q_B \leq$
0.2 AU resulted in planetary systems with $S_s$ values that are
consistent with the $S_s$ distribution of the SJS set.  Note that
one-fourth of the simulations of binary stars
with $Q_B \geq$ 0.3 AU produced only a single terrestrial planet, $N_p$ = 1, in which case
$S_s$ is omitted.  If we consider only SJS and close binary
simulations which resulted in $N_p \geq$ 3 (since two planet systems
generally give high values of this statistic), then the SJS
distribution has an average value of $S_s$ = 38, the same as the Solar
System's terrestrial planets.  Comparing this limited distribution of
$S_s$ to close binary simulations (again omitting 2 planet systems) with
$Q_B \leq$ 0.2 AU (nearly all simulations with larger $Q_B$ resulted
in 1 and 2 planet systems), we again find that only the CB\_.1\_0\_.5 set
results in a statistically different (lower) value of $S_s$ than the SJS $N_p \geq$ 3 distribution.

The angular momentum deficit ($S_d$), which measures the orbital
excitation of a planetary system, is an order of magnitude larger for
the set of planet systems that formed in the SJS simulations ($S_d$ =
0.02, on average) than for the Solar System terrestrial planets ($S_d$
= 0.002).  Most of the planetary systems that formed around close
binaries with $Q_B \leq$ 0.3 AU (in simulations in which the midplane
of the disk began coplanar to the stellar orbit) have values of $S_d$
that are on average comparable to the SJS distribution.  The
exceptions are the two sets of simulations with $a_B$ = 0.1 and $e_B$
= 0, both of which have smaller values of $S_d$ (but even these sets
have higher angular momentum deficit than do the actual terrestrial
planets in our Solar System).  The relatively high values of final
planetary eccentricities and inclinations are a well known difficulty
of models of the late stages of terrestrial planet growth within our
Solar System (Agnor \etal\ 1999, Chambers 2001).

The radial distribution of mass in the final planetary systems is
measured herein by the mass concentration statistic, $S_c$.  In the
Solar System, $\sim$ 90\% of the mass of the terrestrial planets is
concentrated in Venus and in Earth ($S_c$ = 90).  With the exception
of the simulations which resulted in a single planet (for which this
value is infinite), $S_c$ ranged from 21 -- 81 (with an average of
$S_c$ = 40) in the SJS set of runs.  In all of the close binary
simulations that ended with $N_p \geq$ 3, $S_c$ ranged from 25 -- 85.
For most of the two-planet systems, $S_c$ had much higher values,
especially in systems with $Q_B >$ 0.2 AU.  

The degree of radial mixing, $S_r$, which sums the radial migrations
of the bodies that form a final planet, is not known for our Solar
System, but has an average value of $S_r$ = 0.42 for the SJS planetary
systems.  The values of $S_r$ are more widely varied among each close
binary set.  As $Q_B$ is increased and more mass is lost, the degree
of radial mixing is reduced, and the sets of simulations with $Q_B
\geq$ 0.2 AU have smaller values of $S_r$ by a statistically
significant amount than do simulations around a single star.

The percentage of mass that was lost in most of the close binary sets
of simulations is statistically inconsistent with the average total
mass loss in the SJS set of runs ($m_l \sim$ 26\%).  Simulations with
binary stars on circular orbits with $a_B \leq$ 0.1 AU resulted in
planetary systems that accreted, on average, more of the initial disk
mass ($m_l \sim$ 15\% -- 18\%) than the SJS runs, which can be
attributed to both the relatively weak stellar perturbations on the
inner edge of the disk compared to the other binary star systems, and
to the lack of secular resonances from the giant planets near the
outer edge of the disk, which are an important source of perturbations
in the SJS systems.  The amount of mass lost in the close binary
simulations was typically much higher in systems with larger $Q_B$, and only the
CB\_.075\_1/3\_.5 set and the CB\_.2\_0\_.5 set resulted in a
comparable amount of mass loss as the SJS runs, as did the set which
began with the midplane of the disk inclined by 30$^{\circ}$ to the
binary orbital plane, CB\_.1\_0\_.5\_30$^{\circ}$.  More mass was lost
and fewer planets were formed (on average) when the stellar masses
were unequal ($\mu$ = 0.2), since binaries with more extreme mass
ratios (smaller $\mu$) travel farther from the center of mass of the
pair.

In addition to the differences in mass loss, all of the close binary
simulations that began with the disk co-planar to the stellar orbit
resulted (on average) in smaller values of the specific energy
($E/E_{0}$), and greater values of the specific angular momentum
($L/L_{0}$) and specific $Z$-component of angular momentum
($L_Z/L_{Z_0}$).  This is also probably related to the clearing of the
region near 2 AU by the $\nu_{6}$ resonance in the SJS system.  In
close binary systems with $Q_B \geq$ 0.2 AU, the final specific energy
was higher than (or comparable to) the system's initial specific
energy ($E/E_{0} \gtrsim$ 1), although still not as high as the
average value of $E/E_{0}$ for the SJS runs ($\sim$ 1.26).  As $a_B$
and/or $e_B$ is increased, and planets form farther from the binary
center of mass, the specific energy decreases (even to as low as 50\%
of the system's initial specific energy) while the changes in angular
momenta increase by as much as 30\%.

The simulations that began with the midplane of the disk inclined
relative to the stellar orbit, CB\_.1\_0\_.5\_30$^{\circ}$, were 
the only set that resulted in a greater average value of $E/E_{0}$ and
a smaller average $L/L_{0}$ than the SJS runs.  These are both
consequences of the inward drift of the disk resulting from the
initial tilt of the disk to the binary orbital plane.  Even larger
manifestations of this effect are evident in the high-inclination
simulations of terrestrial planet formation within the $\alpha$
Centauri AB system (Quintana \etal\ 2002).  The amount of mass loss
(an average of 25\%), the average specific energy, and the average
specific angular momentum are statistically consistent with the SJS
simulations, but this is simply a result of the coincidental
cancellation of greater inward motion and the absence of the $\nu_{6}$
resonance.  However, the $Z$-component of angular momentum had values
that were smaller than in the SJS simulations, because the final
systems have similar total angular momentum yet larger inclinations
(which also lead to a statistically significant increase of the
angular momentum deficit, $S_d$).

\section{Summary and Conclusions}
In the present work, we have examined the effect of 14 different
short-period binary star configurations (each with a combined stellar
mass of 1 M$_{\odot}$) on the late stages of terrestrial planet
formation within a circumbinary protoplanetary disk.  Stellar mass
ratios of 1:1 and 4:1 were examined, and the initial orbits of the
stars were varied (with semimajor axes between 0.05 AU $\leq a_B \leq$
0.4 AU and eccentricities $e_B \leq$ 0.8) such that the stellar
apastron ranged from 0.05 AU $\leq Q_B \leq$ 0.4 AU.  The midplane of
the disk began coplanar to the stellar orbit in all but one set of
runs; in that exceptional set, the initial inclination of the disk
started at 30$^{\circ}$ relative to the binary orbital plane.  Giant
planets analogous to Jupiter (at $\sim$ 5.2 AU) and, in all but one
set of runs, Saturn (at $\sim$ 9.5 AU) were included.  The evolution
of the protoplanets was followed using a symplectic `close binary'
algorithm which was developed for this purpose (Chambers \etal\ 2002),
and 5 or 6 simulations were performed for each binary star system
under study (with small changes in the initial conditions of the disk)
to account for the chaotic nature of these $N$-body systems.  We
statistically compared our results to a large set of simulations of
the Sun-Jupiter-Saturn system that began with virtually the same
initial disk mass distribution (initially performed by Chambers
(2001), but also integrated herein (Appendix A)).

The close binary stars with maximum separations $Q_B$ $\equiv$ $a_B$(1
+ $e_B$) $\leq$ 0.2 AU and small $e_B$ had little effect on the accreting bodies, and
in most of these simulations terrestrial planets formed over
essentially the entire range of the initial disk mass distribution
(and even beyond 2 AU in many cases).  The stellar perturbations cause
orbits to precess, thereby moving secular resonances out of the inner
asteroid belt, allowing terrestrial planets to form from our initially
compact disk and remain in stable orbits as far as 2.98 AU from
the center of mass of the binary stars.

The effects of the stellar perturbations on the inner edge of the
planetesimal disk become evident in systems with larger $a_B$ (and
$Q_B \geq$ 0.3 AU) and in most of the simulations with $e_B
\textgreater$ 0.  Terrestrial-mass planets can still form around
binary stars with nonzero eccentricity, but the planetary systems tend
to be sparcer and more diverse.  Binary stars with $Q_B \geq$ 0.3 AU
perturb the accreting disk such that the formation of Earth-like
planets near 1 AU is unlikely.  Despite these constraints, at least
one terrestrial planet (at least as massive as the planet Mercury)
formed in each of our simulations.

\acknowledgements
\section*{Acknowledgements}
We thank John Chambers for performing additional simulations presented
in Appendix A.  Fred Adams, John Chambers, and two anonymous referees
provided informative suggestions which helped us improve the
manuscript. This research was supported in part by an Astrobiology
Grant (21-344-53-1y) and a grant from the NASA Ames Director's
Discretionary Fund.  E. V. Q. is a NAS/NRC, and she was supported in
the early stages of this research by a NASA Graduate Student
Researchers Program Fellowship and a Michigan Center for Theoretical
Physics Fellowship.

\appendix
\section{Terrestrial Planet Formation in the Sun-Jupiter-Saturn System}
The initial conditions of the circumbinary disk used in our close
binary simulations are taken from an earlier study of terrestrial
planet formation within a disk around the Sun with Jupiter and Saturn
perturbing the system (Chambers 2001).  In a set of 16 accretion
simulations, Chambers (2001) varied the initial mass distribution of
the circumstellar disk with the intent of determining which conditions
best resulted in the formation of terrestrial planets with masses and
orbits similar to those in the Solar System.  Approximately 150 rocky
bodies (with a total disk mass $\thickapprox$ 2.5 M$_{\oplus}$) were
placed between 0.3 AU -- 2 AU of the Sun, with a surface density that
is consistent with the minimum mass model of the solar nebula
(Weidenschilling 1977).  The masses of the bodies in the disk began
with either a uniform or a `bi-modal' distribution (in which half of
the disk mass is divided into Mars-sized planetary embryos that are
embedded in a swarm of lunar-sized planetesimals).  The radius of each
body was calculated assuming a material density of 3 g cm$^{-3}$.  The
disk was initially placed in the invariable plane of Jupiter and
Saturn, and each system was evolved forward in time for $\sim$ 150 --
300 Myr.

Two simulations, labeled `J21' and `J22' in Chambers (2001), were the
most successful in reproducing terrestrial planets similar to those in
the Solar System.  Each of these began with the bi-modal mass
distribution, and differed only in their randomly chosen initial
arguments of perihelion, longitudes of the ascending node, and mean
anomalies of the embryos and planetesimals.  We chose to use the
initial disk mass distribution from the J21 run for all of our close binary
simulations in order to delineate the effects of an inner binary star
system from the perturbations of a single star (i.e., varying the
initial disk mass distribution in addition to examining the enormous
binary star parameter space is computationally intensive).

To examine the statistical differences between planets formed around
close binaries and those formed around the Sun, we performed an
additional 27 simulations of the Sun-Jupiter-Saturn system (using
conditions nearly identical to the J21 and J22 integrations) to
provide a larger distribution of final planetary systems that form around
the Sun.  Table 3 presents the planetary statistics (as described in
Subsection 4.1) for all of the SJS simulations.  In Group 1, the
original $\it{Mercury}$ hybrid symplectic algorithm (Chambers 1999)
was used.  Group 1a lists the planetary statistics for the original
J21 and J22 simulations of Chambers (2001).  The next two groups each
begin with the same initial conditions as J21 (Group 1b) or J22 (Group
1c), with the exception of a small shift in the initial mean anomaly
of one planetesimal by 1 -- 9 meters.  Only the J22 simulation from
Group 1a was run for 150 Myr, all other simulations of the SJS system
listed into Table 3 were integrated for 200 Myr.

The $\it{Mercury}$ algorithm was recently modified to model planetary
accretion in binary star systems (Chambers \etal\ 2002), and we used
the hybrid symplectic algorithm that is built in this version to
perform the integrations listed in Group 2.  In theory, this hybrid
algorithm should produce results that are consistent with the hybrid
algorithm in the original $\it{Mercury}$ code.  The two simulations in
Group 2a, which had identical starting conditions as those in Group
1a, were performed by J. E. Chambers (private communication, 2004),
and produced results statistically consistent with Group 1a.  The nine
simulations in Group 2b (which began with the same initial conditions
as the corresponding runs from Group 1b) resulted in fewer planets and
more mass loss, on average, than the final systems from Group 1b.

Table 3 gives the average values ($\bar{x}$) for the SJS runs (Group 1
and Group 2), and a statistical comparison (using the
Kolmogorov-Smirnov (K-S) test) between Group 1b and Group 1c, and
between Group 1 and Group 2.  The K-S value, D, and the associated
probability, P, show that the planetary statistics of the final
planets that form in the J21 set of simulations are consistent with
those that formed in the J22 set of runs.  Comparing the distributions
between Group 1 and Group 2 yield different results.  This may be due
to the fact that six of the nine runs from Group 2b resulted in
two-planet systems, which may be a result of small number statistics.
Also note that the simulation which yielded a single planet does not
have a finite value for $S_s$ and $S_c$.  In this case, the highest
value for these statistics is taken when calculating the average, D,
and P, in order to prevent biased results.  We include all of the runs
from Group 1 and Group 2 when comparing the final systems that form
around close binaries (see Table 2).  Finally, Table 3 includes the
planetary statistics for 3 simulations that began with the J21 initial
disk around the Sun, with neither giant planets nor a stellar
companion perturing the system (Quintana \etal\ 2002).  As expected,
the final planetary systems that form in these runs are qualitatively
different than when massive companions are included.

\section{Growth From a Disk of Planetesimals With Forced Eccentricities}
In our nominal circumbinary disk model, the embryos and planetesimals
begin on nearly circular and coplanar orbits about the center of mass
of the pair of stars.  These initial conditions may not be the most
appropriate for simulations in which the binary stars begin with
larger stellar separations and higher eccentricities.  The least
excited state orbiting highly eccentric binary star systems is one in
which the bodies in the disk have higher initial eccentricities and
aligned arguments of periastron.  To examine the magnitude of this
effect on the final outcome of the planets that form, we have
performed an additional set of simulations of the $a_B$ = 0.2 AU and
$e_B$ = 0.5 binary star system (with equal mass stars of 0.5
M$_{\odot}$), altering the initial disk conditions according to the
following prescription.

We first performed a simulation of the $a_B$ = 0.2 AU and $e_B$ = 0.5
system (omitting the giant planets) with massless test particles,
beginning with the coordinates of the planetesimals and planetary
embryos in our nominal disk.  The particles were followed for 100
years (their eccentricity oscillations ranged from periods of years to
decades), and the maximum eccentricity of each test particle,
$e_{max}$, was calculated.

We then set up grids with varying values of the eccentricity
($e$/$e_{max}$) and argument of periastron ($\omega$) for each of the
154 particles, and ran each system for 100 years in order to find the
value of ($e$, $\omega$) that resulted in the smallest range in the
magnitude of the eccentricity.  We varied $e$ and $\omega$ as follows,
while keeping the remaining four orbital elements ($a$, $i$, $\Omega$,
and $M$, which were randomly selected) constant:

$\bullet$ 0.21 $\leq e/e_{max} \leq$ 0.7 (with 0.01 intervals), with
72 values of $\omega$ (at 5$^{\circ}$ intervals) examined for each
$e$/$e_{max}$.

$\bullet$ 0.01 $\leq e/e_{max} \leq$ 0.2 (with 0.01 intervals), with
36 values of $\omega$ (at 10$^{\circ}$ intervals) examined for each $
e/e_{max}$.

Within each grid, the particle which had the smallest range of
eccentricity ($e_{sup}$ - $e_{inf}$) was found, and the initial
coordinates of that particle were noted.  Excluding the values of
semimajor axis at which all bodies were ejected from the system, the
preferred value of $e/e_{max}$ ranged from 0.01 to 0.66, with an
average of 0.096.  We then used these new values for the initial
orbital elements of the bodies in the circumbinary disk, and performed
five simulations with massive planetesimals and embryos, varying the
position of one planetesimal in the disk, as done for our other
planetary growth simulations.  

Figure 9 shows the bodies in the disk at the beginning of the
simulation and also at time $t$ = 200,000 years for the CB\_.2\_.5\_.5
(a -- e) set (left column) and the CBecc\_.2\_.5\_.5 (a -- e) set
(right column).  Even in this early stage of the simulations, the
evolution of the two circumbinary disks are comparable.  Figure 10
presents the final planetary systems for each of these simulations.
The planetary statistics (as described in Section 4.1) for the final
planetary systems are listed in
Table 4 and labelled CBecc\_.2\_.5\_.5 (a -- e).  Table 4 also
re-states the analogous statistics for the five original runs with
$a_B$ = 0.2 AU and $e_B$ = 0.5 that we performed, labelled as
CB\_.2\_.5\_.5 (a -- e).  We used a K-S test to compare planetary
systems resulting from the two sets of simulations, and found that the
distributions for each of the statistics listed are statistically
consistent.  We thus conclude that our choice of initial conditions
for our nominal circumbinary disk are sufficient even for the highly
eccentric binary star systems examined in this article.

One planetesimal in simulation CBecc\_.2\_.5\_.5\_e ended up stranded
in a distant orbit, with periapse well beyond the orbits of the giant
planets (Figure 11).  This planetesimal was scattered outwards by the
giant planets at $t$ = 23.6 yr, and became trapped in a 5:2 resonance
with the outer giant planet at $t$ = 27.8 yr.  This resonance
substantially reduced the planetesimal's eccentrity to $e$ = 0.4,
raising its periapse distance to above the apoapse of the orbit of
the outermost giant planet (the apoapse of this never exceeded $Q_{\saturn}$
= 10.84 AU during the entire integration).  The planetesimal broke free
of the resonance around $t$ = 47.8 yr, but its periapse distance
remained high for the remainder of the simulation.  It is clearly in a
chaotic orbit, capable of returning to the planetary region and being
ejected from the system, but this may well not occur for many
millions, if not billions, of years. Similar processes may have
operated to remove trans-Neptunian objects in our Solar System from
Neptune's immediate control, and if they happened during the migration
epoch of the giant planets, such objects could have been permanently
stranded away from the planetary region (Gomes \etal\ 2005).

\section{Scaling Planetary Accretion Simulations}
The ensemble of plausible initial conditions for simulations of
planetary growth is immense.  Fortunately, one single numerical
simulation (or a set thereof) may be valid for a range of stellar and
planetary masses, orbital periods, etc.  For simulations to correspond
exactly, the following must apply:

\begin{description}
\item[(1)] The ratios of the masses of all of the bodies must be the
  same, i.e., there is a uniform factor by which all masses must be
  multiplied, $M_{\ast}$.

\item[(2)] The ratios of the distances between objects must be
  multipled by an equal amount $r_{\ast}$, and the physical sizes of
  the objects, as well as other factors involving removal of objects
  (inner, outer boundaries in semimajor axis), must be multiplied by
  this same factor $r_{\ast}$.  Note that, in general, the densities
  of objects will change according to the following formula:

\begin{equation}
\rho_{\ast} = \frac{M_{\ast} \;}{ r_{\ast}^3} \; \;. 
\end{equation}

\item[(3)] Initial (and thus subsequent) velocities must be scaled so
that orbital geometries remain the same (``orbital elements'').  For
this to hold, the ratio of kinetic to potential energy must remain the
same, so

\begin{equation}
\frac{M \; v^2}{\frac {M^2} {r}} \; \; = constant \;,
\end{equation}

\begin{equation}
v^2 \propto \frac{M}{r} \; \;. 
\end{equation}

Therefore,

\begin{equation}
v_{\ast} = \left( {\frac{M_{\ast}}{r_{\ast}}} \right) ^ {\frac{1}{2}} \; \;.
\end{equation}
 
\item[(4)] Time must scale in a manner such that the same processes
  take the same number of orbits, so:

\begin{equation}
t_{\ast} = \frac{r_{\ast}}{v_{\ast}} = \left( \frac{
  r_{\ast}^3}{M_{\ast}}\right) ^{\frac{1}{2}} .
\end{equation}

\end{description}

One example of this scaling is that if the star's mass is increased,
and those of the planetesimals/planetary embryos are increased by the
same amount, then the simulations would be applicable either if
planetesimal densities increase by this amount or if distances grow by
the same factor as the physical radii of the planetesimals.  A less
trivial example is that a simulation of the growth of rocky
planetesimals with $\rho$ = 4 g cm$^{-3}$ at 2 AU corresponds to
rock-ice planetesimals with $\rho$ = 1.5 g cm$^{-3}$ at 2.77 AU around
the same single star or around another binary with the same masses but
with a 38\% larger semimajor axis.

Note that we are free to select any positive values for two parameters
($M_{\ast}$ and $r_{\ast}$ in the above discussion), but that these
choices specify the values of the other three parameters.  Collisional
outcomes must also be scaled appropriately, although this is trivially
satisfied by the ``perfect accretion'' assumption used herein.
Additionally, any luminosity-related items, temperature related items,
radiation forces, etc. (none of which were included in our
simulations) must be scaled appropriately.

{}

\begin{deluxetable}{llllllll}
\tablecolumns{7}
\tablecaption{Stability Regions for Test Particles Orbiting Close Binary Stars}
\label{tbl-1}
\tablewidth{0pt}
\tablehead{
\colhead{$\it e_B$} &
\colhead{} &
\colhead{$\it \mu$} &
\colhead{} &
\colhead{} &
\colhead{$a_c$ ($a_B$)} &
\colhead{} &
\colhead{$a^{\ast}_{c}$ ($a_B$)} }

\startdata
0 & & 0.2& & & 2.2& &       1.80  \\
0 & & 0.5& & & 2.3& &       2.01  \\
1/3&& 0.2& & & 3.5 [3.8]&&  3.15  \\
1/3&& 0.5& & & 3.2 [3.5]&&  2.99  \\
0.5&& 0.2& & & 3.9& &       3.40  \\
0.5&& 0.5& & & 3.6& &       2.98  \\
0.8&& 0.2& & & 4.3& &       3.86  \\
0.8&& 0.5& & & 3.9& &       3.19  \\
\enddata
\end{deluxetable}

\begin{deluxetable}{lllllllllllllll}
\tablecolumns{15} 
\tabletypesize{\scriptsize} 
\rotate
\tablecaption{Statistics for Final Planetary Systems}
\label{tbl-2}
\tablewidth{0pt} 
\tablehead{
\colhead{$\bf Run$} &
\colhead{$\it t$ (Myr)} &
\colhead{\bf $N_p$}   &
\colhead{\bf $N_m$}   &
\colhead{\bf $a_p$/$a^{\ast}_{c}$}   &
\colhead{$q_p$/$Q_{B}$}     & 
\colhead{$S_m$}     & 
\colhead{$S_s$}     & 
\colhead{\bf $S_d$} & 
\colhead{\bf $S_c$} & 
\colhead{\bf $S_r$} &
\colhead{\bf $m_l$ (\%)} &
\colhead{\bf $E/E_{0}$} & 
\colhead{$L/L_{0}$} & 
\colhead{\bf $L_Z/L_{Z_0}$}}
\startdata 
&&&&&&&&&&&&&&\\    
CB\_.05\_0\_.5$^\dag$\_a& 500   &6&1&3.53&7.12&0.288&27.1&0.0041&32.3&0.226&15.71&1.13&0.94&0.93\\
CB\_.05\_0\_.5$^\dag$\_b& 200   &4&1&3.48&7.18&0.388&36.3&0.0063&29.5&0.343&13.57&1.14&0.93&0.93\\
CB\_.05\_0\_.5$^\dag$\_c& 500&4&1& 3.48&7.28& 0.515&34.4&0.0008&36.8&0.289&14.29 &1.13&0.93&0.93\\   
CB\_.05\_0\_.5$^\dag$\_d &200  &3&1&5.00&10.47&0.458&43.0&0.0047&34.7&0.410&18.93&1.18&0.91&0.91\\
CB\_.05\_0\_.5$^\dag$\_e  &200    &2&1&7.47&16.02&0.905&40.0&0.0110&213.8&0.437&28.93&1.33&0.82&0.81\\   
\bf{$\bar{x}$}&320 &3.8&  1   &    4.59&   9.61&   0.511&  36.2&   0.0054& 69.4&   0.341&18.29 & 1.18& 0.91& 0.90\\
D &$\cdots$&0.28&0.58 & $\cdots$& $\cdots$&0.22&0.28&0.48&0.35&0.28&0.70&0.67&0.67&0.70\\  
P &$\cdots$&0.832&0.066 & $\cdots$& $\cdots$&0.968&0.832&0.200&0.572&0.810&0.014&0.021&0.021&0.014\\  

&&&&&&&&&&&&&&\\   
CB\_.05\_0\_.5\_a   &500  &3&1&3.42&7.52&0.500&45.3&0.0069&30.8&0.379&15.71&1.15&0.93&0.93\\
CB\_.05\_0\_.5\_b   &200     &3&1&3.46&7.29&0.561&48.4&0.0252&29.5&0.454&24.29&1.23&0.88&0.88\\
CB\_.05\_0\_.5\_c    &500       &5&1&5.00&10.95&0.355&30.3&0.0026&31.4&0.309&11.43&1.11&0.95&0.94\\
CB\_.05\_0\_.5\_d   &200         &3&1& 5.12&10.50&0.473&42.4&0.0049&29.7&0.364&12.50&1.13&0.94&0.94\\
CB\_.05\_0\_.5\_e &500  &5&1&3.43&7.17&0.294&27.8&0.0066&27.7&0.204&11.43&1.10&0.96&0.95\\   
\bf{$\bar{x}$}&380&3.8 &1& 4.09& 8.68& 0.437&  38.84 &  0.0092& 29.8 &0.342&15.07& 1.14& 0.93&      0.93\\  
D    & $\cdots$&0.37&0.58&$\cdots$&$\cdots$&0.34&0.37&0.41&0.67&0.38&0.80&0.80&0.77&0.77\\   
P  &$\cdots$&0.501&0.066&$\cdots$&$\cdots$&0.620&0.501&0.355&0.021&0.457&0.003&0.003&0.005&0.005\\   

&&&&&&&&&&&&&&\\
CB\_.075\_1/3\_.5\_a     &200    &3&1&1.56&3.66&0.436&40.2&0.0166&27.7&0.363&24.64&1.02&1.00&0.98\\
CB\_.075\_1/3\_.5\_b    &200    &2&0&5.89&16.47&0.910&54.1&0.0310&167.7&0.290&52.50&0.75&1.06&1.06\\
CB\_.075\_1/3\_.5\_c    &200    &3&0&2.21&5.96&0.566&41.3&0.0047&37.5&0.306&13.57&1.08&0.95&0.95  \\
CB\_.075\_1/3\_.5\_d   &200      &2&3&5.50&13.39&0.487&42.9&0.0282&96.2&0.285&59.64  &0.68&1.13&1.13\\
CB\_.075\_1/3\_.5\_e   &200   &3&1&1.54&3.58&0.489&42.9&0.0092&35.3&0.255&21.79&1.15&0.92&0.92\\   
\bf{$\bar{x}$}&200 &2.6   &  1 &      3.34&  8.61& 0.578&  44.28 &  0.0179& 72.9 &  0.300&  34.43&  0.94&   1.01&   1.01\\   
D   &$\cdots$  &0.48&0.34 &$\cdots$&$\cdots$ &0.42&0.55&0.37&0.37&0.64&0.37&0.80&0.935&0.968\\   
P  &$\cdots$&0.200&0.620 &$\cdots$&$\cdots$ &0.337&0.096&0.479&0.501&0.032&0.501&0.003&0.000&0.000\\   

&&&&&&&&&&&&&&\\    
CB\_.1\_0\_.5\_a     &200       &4&1&2.40&5.70&0.478&36.1&0.0033&32.7&0.305&17.14&1.11&0.94&0.94\\
CB\_.1\_0\_.5\_b      &200        &5&1&1.80&3.74&0.313&29.8&0.0017&29.7&0.314 &11.06&1.07&0.97&0.97\\
CB\_.1\_0\_.5\_c     &200       &4&1&1.88&4.22&0.335&33.2&0.0040&32.9&0.253&11.43&1.06&0.97&0.97\\
CB\_.1\_0\_.5\_d  &200  &4&1&2.97&6.44&0.297&30.7&0.0044&35.9&0.263&17.14&1.06&0.96&0.96\\
CB\_.1\_0\_.5\_e  &200   &4&1&2.62&5.84&0.345&30.1&0.0036&39.0&0.314&21.43&1.19&0.90&0.90\\ 
\bf{$\bar{x}$}&200&4.2 &    1&       2.33& 5.19 & 0.354&  32.0 &  0.0034&       34.0  & 0.290&15.64& 1.10& 0.95&   0.95\\ 
D    & $\cdots$&0.68&0.58&$\cdots$&$\cdots$&0.74&0.74&0.70&0.47&0.57&0.77&0.80&0.77&0.77\\   
P& $\cdots$ &0.019&0.066&$\cdots$&$\cdots$&0.008&0.008&0.014&0.212&0.071&0.005&0.003&0.005&0.005\\   

&&&&&&&&&&&&&\\    
CB\_.1\_0\_.2\_a  &200    &3&1&2.00&3.85&0.440&41.1&0.0038&35.6&0.314&19.29&1.11&0.94&0.94\\
CB\_.1\_0\_.2\_b  &200    &4&1&2.63&5.45&0.322&34.9&0.0053&29.7&0.422&16.79&1.14&0.94&0.94\\
CB\_.1\_0\_.2\_c  &200    &4&0&3.04&5.86&0.451&33.6&0.0019&33.4&0.317&11.79&1.07&0.96&0.96\\  
CB\_.1\_0\_.2\_d  &500&4&1& 1.95 &3.73  &0.500&29.7&0.0041&36.9&0.393&20.00 &1.17&0.91&0.91\\ 
CB\_.1\_0\_.2\_e  &500&5&0&2.56&5.00&0.414&32.5&0.0031&31.2&0.305&18.93&1.13&0.94&0.94\\ 
\bf{$\bar{x}$}&320&4.0&0.6&     2.44& 4.78&  0.425&  34.36&   0.0036&       33.4&  0.350&    17.36& 1.12& 0.94&        0.94\\
D&$\cdots$&0.48&0.34& $\cdots$& $\cdots$&0.38&0.54&0.67&0.45&0.37&0.67&0.80&0.87&0.90\\    
P&$\cdots$&0.200&0.620& $\cdots$& $\cdots$&0.457&0.103&0.021&0.269&0.479&0.021&0.003&0.001&0.001\\    

&&&&&&&&&&&&&&\\
CB\_.1\_.8\_.5\_a  &200   &3&2&2.29&4.31&0.440&35.3&0.0284&61.2&0.273&31.79&0.90&1.01&0.99\\
CB\_.1\_.8\_.5\_b &500&2&0& 2.45 &4.74&0.667&45.3&0.0100&78.8&0.305&38.93&0.95&0.98&0.98\\    
CB\_.1\_.8\_.5\_c  &200    &3&2&1.84&3.33&0.634&43.4&0.0101&55.9&0.219&31.79&1.00&0.97&0.96\\  
CB\_.1\_.8\_.5\_d &200     &3&2&1.90& 3.82&0.574&49.0&0.0276&37.2&  0.274&39.64&1.06&0.95&0.93\\
CB\_.1\_.8\_.5\_e &200    &4&1&1.52&2.88&0.328&36.5&0.0451&33.9&0.291&30.36&1.09&0.95&0.91\\ 
\bf{$\bar{x}$} & 650&3.0&1.4&2.00&3.82& 0.529&  41.9&    0.0242&   53.4 &   0.272& 34.50 & 1.00& 0.97&  0.96\\
D&$\cdots$&0.45&0.41& $\cdots$& $\cdots$&0.37&0.26&0.52&0.50&0.67&  0.77  &0.80&0.97&0.87\\    
P&$\cdots$&0.269&0.374& $\cdots$& $\cdots$&0.479&0.889&0.135&0.155&0.021&0.005  &0.003&0.000&0.001\\    

&&&&&&&&&&&&&&\\
CB\_.1\_.8\_.2\_a  &500  &2&0&1.98&4.86&0.544&83.5&0.0841&19.7&0.527& 46.79&0.95&1.03&0.97\\
CB\_.1\_.8\_.2\_b  &500    &3&0&1.93&4.50&0.468&38.7&0.0276&57.5&0.347&32.86&1.09&0.92&0.90\\	
CB\_.1\_.8\_.2\_c  &200    &3&1&2.13&4.91&0.433&43.5&0.0060&39.5&0.208&35.71&0.93&1.02&1.02\\
CB\_.1\_.8\_.2\_d &165   &1&0&2.82&6.78&1.000&$\cdots$&0.0109&$\infty$&0.376&46.43&0.95&0.95&0.95\\
CB\_.1\_.8\_.2\_e  &200    &3&1&1.52&3.96&0.490&47.6&0.0378&30.5&0.368&30.00&1.02&0.98&0.95\\ 
\bf{$\bar{x}$}&313& 2.4 &    0.4    & 2.07& 5.00& 0.587&   59.4& 0.0333&       40.9&   0.365& 38.36& 0.99&        0.98& 0.96\\
D   &  $\cdots$&0.48&0.50&$\cdots$&$\cdots$&0.39&0.52&0.47&0.30&0.38&0.71&0.80&0.94&0.77\\   
P &$\cdots$ &0.200&0.155&$\cdots$&$\cdots$&0.435&0.135&0.212&0.742&0.457&0.012&0.003&0.000&0.005\\   

&&&&&&&&&&&&&&\\
CB\_.1\_0\_.5\_30$^{\circ}$\_a  &500 &4&0&2.47&5.46&0.453&38.5&0.1182&29.8&0.387 & 28.21&1.42&0.83&0.74\\ 
CB\_.1\_0\_.5\_30$^{\circ}$\_b &500  &2&1&2.18&5.06&0.724&50.5&0.0938&60.1&0.513&22.50&1.53&0.78&0.71\\ 
CB\_.1\_0\_.5\_30$^{\circ}$\_c  &500 &3&0&1.98&4.29&0.515&48.1&0.1473&25.5&0.467&18.93&1.48&0.82&0.71\\
CB\_.1\_0\_.5\_30$^{\circ}$\_d &200  &4&1& 1.79&3.69&0.421&36.5&0.0157&26.7&0.293&18.57&1.12&0.95&0.94\\
CB\_.1\_0\_.5\_30$^{\circ}$\_e &550&2&0&4.07&9.50&0.913&49.9&0.1610&186.1&0.361&42.14&1.21&0.85&0.72\\
\bf{$\bar{x}$}&450&3.0      & 0.4&     2.50& 5.60&     0.605&  44.7 &   0.1072&        65.6&   0.404&       26.07& 1.35&  0.85&        0.76\\
D &$\cdots$&0.31&0.50&$\cdots$&$\cdots$&0.39&0.32&0.77&0.47&0.19&0.30&0.54&0.57&0.77\\
P &$\cdots$&0.718&0.155&$\cdots$&$\cdots$&0.435&0.669&0.005&0.212&0.994&0.742&0.110&0.071&0.005\\

&&&&&&&&&&&&&&\\
CB\_.15\_1/3\_.5\_a    &200  &2&0&2.03&5.16&0.736&52.4&0.0203&79.5&0.174&56.79&0.77&1.09&1.08\\
CB\_.15\_1/3\_.5\_b   &200   &3&3&1.63&3.91&0.600&33.6&0.0070&83.7&0.313&44.64&1.09&0.92&0.91\\
CB\_.15\_1/3\_.5\_c    &200   &2&0&2.40&6.19&0.714&39.5&0.0135&120.6&0.275&45.00&0.86&1.02&1.01\\
CB\_.15\_1/3\_.5\_d  &200      &3&1& 1.74&4.10&0.626&44.3&0.0134&60.4&0.270&37.86&0.93&1.00&0.99\\
CB\_.15\_1/3\_.5\_e  &200  &3&2&1.27&3.10&0.497&39.9&0.0066&57.7&0.260&39.64&1.00&0.97&0.96\\ 
\bf{$\bar{x}$}&200& 2.6&     1.2&     1.82& 4.49& 0.635&  41.9&   0.0122&       80.4&   0.258&        44.79 &0.93&       1.00&        0.99\\
D    &$\cdots$ &0.48&0.34&$\cdots$&$\cdots$&0.61&0.32&0.39&0.90&0.74&0.97&0.80&0.94&0.97\\   
P &$\cdots$ &0.200&0.620&$\cdots$&$\cdots$&0.049&0.693&0.435&0.001&0.008&0.000&0.003&0.000&0.000\\   

&&&&&&&&&&&&&&\\
CB\_.2\_0\_.5\_a   &200       &3&1&1.11&2.35&0.545&47.6&0.0064&40.0&0.182&23.93&1.02&0.97&0.97\\
CB\_.2\_0\_.5\_b    &200      &5&1&1.15&2.56&0.429&29.6&0.0058&45.8&0.248&20.00&1.08&0.94&0.94\\
CB\_.2\_0\_.5\_c    &200      &4&2&1.17&2.42&0.519&32.7&0.0075&58.4&0.405&32.50&1.12&0.91&0.91\\
CB\_.2\_0\_.5\_d     &500   &4&1& 1.07&2.21 &0.553&38.8&0.0053&34.2&0.298&32.14&1.12&0.94&0.93\\
CB\_.2\_0\_.5\_e    &200    &2&1&2.35&4.85&0.633&36.1&0.0052&109.4&0.242&43.57&0.95&0.98&0.97\\ 
\bf{$\bar{x}$}&260&3.6&     1.2 &    1.37& 2.88& 0.536&  37.0&   0.0060&       57.6&  0.275&    30.43& 1.06& 0.95&  0.94\\
D    & $\cdots$&0.28&0.58&$\cdots$&$\cdots$&0.41&0.34&0.45&0.41&0.64&0.44&0.80&0.87&0.87\\   
P  &$\cdots$&0.832&0.066&$\cdots$&$\cdots$&0.35&0.620&0.269&0.35&0.032&0.285&0.003&0.001&0.001\\   

&&&&&&&&&&&&&&\\
CB\_.2\_0\_.2\_a&200 &3&2&1.24&2.30&0.659&59.0&0.0067&41.5&0.263&41.43&0.92&1.02&1.02\\
CB\_.2\_0\_.2\_b&200 &3&0&1.29&2.35&0.632&41.8&0.0041&84.2&0.253&48.57&1.08&0.93&0.92\\  
CB\_.2\_0\_.2\_c&500 &3&3&1.28 &2.51&0.661&48.1&0.0096&62.1&0.233&41.07&1.07&0.93&0.93\\
CB\_.2\_0\_.2\_d&200 &3&0&2.40&4.99&0.461&35.9&0.0073&60.4&0.220&45.00&0.85&1.05&1.05\\  
CB\_.2\_0\_.2\_e&200 &3&0&1.27&2.43&0.487&49.0&0.0058&41.0&0.323&30.36&1.02&0.98&0.97\\  
\bf{$\bar{x}$}&260&3.0&1&1.50&2.92&0.580&46.8&0.0067& 57.8&   0.258& 41.29&      0.99& 0.98& 0.98\\
D     &$\cdots$&0.41&0.50&$\cdots$&$\cdots$&0.42&0.38&0.38&0.68&0.74& 0.77&0.80&0.97&0.97\\
P     &$\cdots$&0.355&0.155&$\cdots$&$\cdots$&0.337&0.457&0.457&0.019&0.008& 0.005&0.003&0.000&0.000\\

&&&&&&&&&&&&&&\\
CB\_.2\_.5\_.5\_a &146     &1&0&2.35&5.44&1.000&$\cdots$&0.0193&$\infty$&0.147&73.58&0.74&1.07&1.07\\
CB\_.2\_.5\_.5\_b &200     &2&0&2.25&4.65&0.890&57.2&0.0021&152.4&0.169&67.50&0.73&1.11&1.11  \\
CB\_.2\_.5\_.5\_c&154     &1&0&3.56&8.54&1.000&$\cdots$&0.0328&$\infty$&0.378&85.71&0.49&1.31&1.30\\
CB\_.2\_.5\_.5\_d&115     &1&0& 2.70&5.92&1.000&$\cdots$&0.0190&$\infty$&0.263&79.29&0.64&1.16&1.15\\
CB\_.2\_.5\_.5\_e &200  &2&0&1.78&5.20&0.590&32.4&0.0104&185.7&0.155&62.50&0.67&1.15&1.15\\ 
\bf{$\bar{x}$}&163&1.4 &0.0& 2.53& 5.95& 0.896&   52.2& 0.0167&179.0&  0.222&73.72 &  0.65&  1.16& 1.15\\
D        & $\cdots$ &0.74&0.67&$\cdots$&$\cdots$&0.77&0.70&0.41&0.97&0.74&0.97&0.80&0.97&0.97\\
P       &$\cdots$ &0.008&0.021&$\cdots$&$\cdots$&0.005&0.014&0.374&0.000&0.008&0.000&0.003&0.000&0.000\\

&&&&&&&&&&&&&&\\
CB\_.2\_.5\_.2\_a  &200     &2&1&1.15&2.95&0.547&36.9&0.0081&133.3&0.127&69.29&0.66&1.17&1.16\\
CB\_.2\_.5\_.2\_b  &200     &2&0&1.85&4.45&0.563&30.9&0.0126&203.1&0.179&63.21&0.69&1.13&1.12\\
CB\_.2\_.5\_.2\_c &500&2&1&1.82&4.62&0.691&67.6&0.0080&71.0&0.425&80.36&0.72&1.13&1.12\\
CB\_.2\_.5\_.2\_d   &500    &2&0&1.39&3.21&0.828&54.2&0.0244&141.1&0.083&77.15&0.73&1.09&1.09\\
CB\_.2\_.5\_.2\_e  &200    &2&1&1.36&3.34&0.860&46.5&0.0099&190.0&0.184&66.79&0.75&1.09&1.09\\ 
\bf{$\bar{x}$}&320&2.0&       0.6  &   1.51&   3.71&   0.698&  47.2&   0.0126&       147.7 &  0.200& 71.36& 0.71& 1.12&  1.12\\
D&$\cdots$&0.64&0.34&$\cdots$&$\cdots$&0.71&0.31&0.45&0.97&0.80&0.97&0.80&0.97&0.97\\
P&$\cdots$&0.032&0.620&$\cdots$&$\cdots$&0.012&0.718&0.254&0.000&0.003&0.000&0.003&0.000&0.000\\

&&&&&&&&&&&&&&\\
CB\_.3\_0\_.5\_a&200 &2&0&1.99&4.63&0.670&37.6&0.0088&153.9&0.166&64.28&0.67&1.15&1.15\\
CB\_.3\_0\_.5\_b&200 &3&0&1.32&2.90&0.667&41.8&0.0044&84.9&0.198&48.58&0.78&1.09&1.09\\
CB\_.3\_0\_.5\_c&200 &2&1&1.75&3.75&0.556&36.2&0.0056&126.2&0.176&55.00&0.80&1.07&1.06\\ 
CB\_.3\_0\_.5\_d&200 &2&0&2.28&4.96&0.734&51.8&0.0141&101.6&0.221&71.79&0.67&1.15&1.15\\
CB\_.3\_0\_.5\_e&200 &2&2&1.51&3.22&0.653&54.2&0.0051&65.9&0.197&56.79&0.80&1.08&1.08\\
\bf{$\bar{x}$}&200&2.2&0.6&1.77&3.89& 0.656&   44.3&  0.0076& 106.5& 0.192& 59.29&0.74& 1.11&1.11\\
D &$\cdots$&0.64&0.50&$\cdots$&$\cdots$&0.74&0.32&0.34&0.97&0.80&0.97&0.80&0.97&0.97\\
P&$\cdots$ &0.032&0.155&$\cdots$&$\cdots$&0.008&0.669&0.596&0.000&0.003&0.000&0.003&0.000&0.000\\

&&&&&&&&&&&&&&\\
CB\_.3\_1/3\_.5\_a  &200   &2&0&1.51&3.91&0.725&40.4&0.0043&235.5&0.245&85.71&0.61&1.21&1.21\\
CB\_.3\_1/3\_.5\_b &200    &2&0&1.41&3.42&0.843&23.3&0.0017&811.3&0.185&75.00&0.68&1.14&1.14\\
CB\_.3\_1/3\_.5\_c &175    &1&0&1.79&4.12&1.000&$\cdots$&0.0011&$\infty$&0.181&86.07&0.64&1.17&1.17\\
CB\_.3\_1/3\_.5\_d&200    &1&1&2.17&4.87&0.960&$\cdots$&0.0053&1839.4&0.167   &91.07 &0.53&1.29&1.29\\
CB\_.3\_1/3\_.5\_e  &100  &1&0&1.75&4.56&1.000&$\cdots$&0.0040&$\infty$&0.105&78.57&0.62&1.19&1.19\\ 
CB\_.3\_1/3\_.5\_f &200   &1&2&1.63&3.96&0.957&$\cdots$&0.0069&541.8&0.208&83.57&0.65&1.17&1.16\\ 
\bf{$\bar{x}$}&179&1.3& 0.5&     1.71& 4.14& 0.914&       37.6&0.0039&      1184.5& 0.182& 83.33 & 0.62&1.19& 1.19\\
D &$\cdots$&0.77&0.54&$\cdots$&$\cdots$&0.90&0.38&0.58&0.97&0.83&0.97&0.83&0.97&0.97\\
P&$\cdots$ &0.002&0.069&$\cdots$&$\cdots$&0.000&0.365&0.043&0.000&0.001&0.000&0.001&0.000&0.000\\

&&&&&&&&&&&&&&\\
CB\_.4\_0\_.5\_a &500  &2&1&1.56&3.24&0.598&34.1&0.0011&153.3&0.206&67.15&0.67&1.16&1.16	\\
CB\_.4\_0\_.5\_b &1000  &1&2&1.34&2.73&0.971&$\cdots$&0.0006&2183.7&0.141&75.00&0.72&1.11&1.11	\\
CB\_.4\_0\_.5\_c &929  &1&0&2.27&5.35&1.000&$\cdots$&0.0214&$\infty$&0.081&90.00&0.57&1.23&1.22	\\
CB\_.4\_0\_.5\_d  &500 &2&2&1.10&2.40&0.595&25.6&0.0024&193.0&0.281&85.00&0.71&1.12&1.12\\
CB\_.4\_0\_.5\_e  &200  &1&1&1.88&4.47&0.986&$\cdots$&0.0008&3356.0&0.282&75.00&0.60&1.22&1.22\\ 
CB\_.4\_0\_.5\_f &129   &1&0&1.99&5.46&1.000&$\cdots$&0.0193&$\infty$&0.222&85.35&0.57&1.22&1.22\\ 
\bf{$\bar{x}$}&543&1.3& 1.0&       1.69& 3.94& 0.858&       32.7&0.0076&     2099.7& 0.202&  79.58& 0.64&1.18& 1.17\\
D &$\cdots$&0.77&0.27&$\cdots$&$\cdots$&0.81&0.64&0.67&0.97&0.74&0.97&0.83&0.97&0.97\\
P &$\cdots$&0.002&0.795&$\cdots$&$\cdots$&0.001&0.017&0.011&0.000&0.004&0.000&0.001&0.000&0.000\\

&&&&&&&&&&&&&&\\    
MVEM    &$\cdots$  &4.0&0.0&$\cdots$&$\cdots$&0.509&37.7&0.0018&89.9&$\cdots$&$\cdots$&$\cdots$&$\cdots$&$\cdots$\\    
SJS\_{ave}& 200&3.0&0.7&$\cdots$&$\cdots$&0.514&44.4&0.0157&40.5&0.420&25.92&1.26&0.87&0.87\\    
Sun\_{ave}  & 867                      &4.0&12.3&$\cdots$&$\cdots$&0.393&38.2&0.0318&15.9&0.360&0.48&1.06&1.01&1.00\\    
Sun\_{ave} ($a <$ 2 AU) &  867 &3.0&0.7&$\cdots$&$\cdots$&0.483&36.8&0.0061&34.3&$\cdots$&18.66$^\ddag$ &1.21&0.90&0.90\\    
$\alpha$ Cen A ($i \leq$ 30$^{\circ}$)&481&3.9&0.2&$\cdots$&$\cdots$&0.432&38.4&0.0098&30.9&0.354&15.69&1.24&0.90&0.94\\    

\enddata 
\tablenotetext{\dag}{Systems include ``Jupiter'' but not ``Saturn''.  All other simulations include both giant planets.}
\tablenotetext{\ddag}{Mass lost plus mass ending up in planets/minor planets beyond 2 AU.}
\end{deluxetable} 

\begin{deluxetable}{lllllllllllll}
\tablecolumns{13} 
\tabletypesize{\scriptsize} 
\rotate
\tablecaption{Sun-Jupiter-Saturn Simulations}
\tablewidth{0pt} 
\tablehead{
\colhead{$\bf Run$} &
\colhead{\bf $N_p$}   &
\colhead{\bf $N_m$}   &
\colhead{$S_m$}     & 
\colhead{$S_s$}     & 
\colhead{$S_d$}     & 
\colhead{\bf $S_c$} & 
\colhead{\bf $S_r$} &
\colhead{\bf $m_{l_\star}$ (\%)} &
\colhead{\bf $m_{l_\infty}$ (\%)} & 
\colhead{\bf $E/E_{0}$} & 
\colhead{$L/L_{0}$} &
\colhead{$Lz/L_{0}$}}
\startdata 

\bf{Group 1a}&&&&&&&&&&&& \\
J21& 4 & 0 & 0.335 & 31.3 & 0.0061 & 37.3 &0.328 &17.14 &0.71 & 1.17 & 0.91 & 0.91 \\
J22& 4 & 0 & 0.326 & 33.7 & 0.0105 & 35.0 &0.281 &18.57 &1.43 & 1.19 & 0.91 & 0.90 \\

\bf{Group 1b}&&&&&&&&&&&& \\
J21\_1    & 3 & 2 & 0.532 & 43.1 & 0.0048 & 39.6 &0.402 &21.43&5.36  & 1.29 & 0.87 & 0.87 \\
J21\_2    & 3 & 1 & 0.379 & 33.3 & 0.0041 & 49.3 &0.508 &18.93&0.00  & 1.24 & 0.87 & 0.87 \\
J21\_3    & 3 & 1 & 0.394 & 36.9 & 0.0032 & 34.9 &0.518 &20.71&0.36  & 1.16 & 0.92 & 0.92 \\
J21\_4    & 3 & 4 & 0.400 & 29.7 & 0.0132 & 61.5 &0.339 &26.79&7.14  & 1.33 & 0.84 & 0.83 \\
J21\_5    & 4 & 0 & 0.496 & 34.0 & 0.0061 & 36.3 &0.362 &20.00&0.00  & 1.17 & 0.92 & 0.91 \\
J21\_6    & 4 & 2 & 0.551 & 35.8 & 0.0083 & 34.8 &0.296 &23.57&0.00  & 1.23 & 0.89 & 0.89 \\
J21\_7    & 2 & 0 & 0.571 & 55.0 & 0.0118 & 40.0 &0.396 &14.64&2.14  & 1.20 & 0.89 & 0.89 \\ 
J21\_8    & 1 & 0 & 1.000 & $\cdots$  & 0.0686 & $\infty$ &0.498 &34.29&1.79  & 1.48 & 0.73 & 0.72 \\
J21\_9    & 3 & 0 & 0.491 & 52.8 & 0.0044 & 22.2 &0.411 &22.50&1.79  & 1.26 & 0.91 & 0.91 \\

\bf{Group 1c}&&&&&&&&&&&& \\
J22\_1   & 3 & 3 & 0.404 & 36.5 & 0.0044 & 34.6 & 0.414 &20.36 &0.00 & 1.20 & 0.91 & 0.90 \\
J22\_2    & 3 & 0 & 0.518 & 40.9 & 0.0117 & 48.7 &0.344 &30.36 &1.43 & 1.33 & 0.84 & 0.84 \\
J22\_3    & 3 & 0 & 0.521 & 46.0 & 0.0188 & 37.5 &0.580 &22.86 &0.36 & 1.24 & 0.88 & 0.87 \\
J22\_4    & 4 & 0 & 0.418 & 33.4 & 0.0088 & 36.1 &0.255 &18.93 &0.71 & 1.18 & 0.91 & 0.91 \\
J22\_5    & 3 & 1 & 0.392 & 34.7 & 0.0069 & 42.3 &0.463 &29.64 &1.07 & 1.31 & 0.86 & 0.85 \\
J22\_6    & 4 & 0 & 0.377 & 36.0 & 0.0033 & 34.1 &0.339 &18.93 & 1.43 &1.18 & 0.91 & 0.91 \\
J22\_7    & 4 & 0 & 0.433 & 37.5 & 0.0066 & 33.1 & 0.311&25.00 &0.71 & 1.25 & 0.89 & 0.88 \\
J22\_8    & 3 & 1 & 0.381 & 45.8 & 0.0260 & 25.0 &0.486 &18.93 &0.36 & 1.19 & 0.92 & 0.91 \\
J22\_9    & 4 & 0 & 0.379 & 36.8 & 0.0063 & 40.2 &0.257 &25.36 &1.43 & 1.21 & 0.89 & 0.89 \\ 

\bf{$\bar{x}$ (Group 1)}&3.25 &0.75 &0.465&39.4 &0.0117&39.2&0.389&22.45&1.41&1.24&0.88&0.88 \\

&&&&&&&&&&&& \\
\bf{Group 2a}&&&&&&&&&&&& \\
J21$^\ast$& 4 & 1 & 0.488 & 34.5 & 0.0122 & 33.3 &0.307  &22.50 &1.43 &1.21&0.90&0.90 \\
J22$^\ast$& 3 & 1 & 0.733 & 42.1 & 0.0077 & 81.0 &0.412  &26.43 &0.00 &1.26&0.86&0.85 \\

\bf{Group 2b}&&&&&&&&&&&& \\
J21\_1$^\ast$&2&0&0.871&87.0&0.0599 & 37.4&0.784 &29.29&12.50&1.49&0.77&0.76\\  
J21\_2$^\ast$&2&0&0.513&65.9&0.0674 & 29.0&0.580&29.29&1.07&1.31&0.83&0.82\\    
J21\_3$^\ast$&2&0&0.651&54.2&0.0259 & 49.0&0.386&25.71&4.64&1.29&0.84&0.84\\
J21\_4$^\ast$&2&0&0.684&80.8&0.0336 & 21.4&0.518&32.50&0.71&1.36&0.86&0.85\\
J21\_5$^\ast$&3&0&0.527&39.8&0.0051 & 51.2&0.465&32.14&0.71&1.33&0.84&0.84\\
J21\_6$^\ast$&2&1&0.592&54.6&0.0036 & 44.0&0.550&26.43&1.79&1.30&0.86&0.86\\
J21\_7$^\ast$&2&0&0.684&54.3&0.0115 & 50.0&0.594&25.71&0.71&1.27&0.86&0.86\\
J21\_8$^\ast$&4&2&0.397&37.2&0.0126 & 35.9&0.278&16.07&9.29&1.25&0.88&0.88\\
J21\_9$^\ast$&3&3&0.488&38.3&0.0122 & 39.2&0.345&26.79&0.71&1.27&0.87&0.87\\

\bf{$\bar{x}$ (Group 2)}&2.64&0.73&0.603&  53.5& 0.0229& 42.9& 0.474&  26.62& 3.05& 1.30& 0.85&       0.85\\
&&&&&&&&&&&& \\

\bf{$\bar{x}$ (All)}&3.03&0.74&0.514&44.4&0.0157&40.5&0.420&23.93&1.99&1.26&0.87&0.87\\

&&&&&&&&&&&& \\

\bf{Group 1b, Group 1c}&&&&&&&&&&&& \\
D     &0.22&0.22&0.44&0.22&0.33&0.33&0.33&0.22&0.44&0.22&0.22&0.22 \\
P  &0.958&0.958&0.250&0.958&0.603&0.603&0.603&0.958&0.250&0.958&0.958&0.958 \\

&&&&&&&&&&&& \\
\bf{Group 1, Group 2}&&&&&&&&&&&& \\
D     &0.57&0.25&0.51&0.51&0.44&0.25&0.40&0.62&0.26&0.56&0.57&0.57 \\
P  &0.012&0.722&0.032&0.032&0.096&0.679&0.147&0.005&0.658&0.014&0.012&0.012 \\

&&&&&&&&&&&& \\
&&&&&&&&&&&& \\
\bf{Sun-only}&&&&&&&&&&&& \\
S\_1&4&9&0.462&39.7&0.0136&19.0&0.335&0.000&0.357&1.06&1.00&1.00 \\
S\_2&4&8&0.357&38.2&0.0342&16.0&0.359&0.000&1.071&1.05&1.00&0.99 \\
S\_3&4&20&0.361&36.7&0.0476&12.8&0.385&0.000&0.000&1.07&1.02&1.00 \\
\bf{$\bar{x}$}&4&12.3&0.393 &38.2 &0.0318 &15.9 &0.360 &0.000 &0.476 &1.06 &1.01 &1.00 \\
&&&&&&&&&&&& \\

\bf{Sun-only, All}&&&&&&&&&&&& \\
D     &0.68&0.97&0.52&0.39&0.77&0.75&0.36&0.75&0.49&0.75&0.97&0.97 \\
P  &0.041&0.001&0.194&0.548&0.012&0.017&0.631&0.017&0.256&0.017&0.001&0.001 \\

&&&&&&&&&&&& \\
\enddata
\tablenotetext{\ast}{Simulations performed using a symplectic hybrid integrator that is incorporated
  in the close binary code.}
\end{deluxetable}

\begin{deluxetable}{lllllllllllllll}
\tablecolumns{15} 
\tabletypesize{\scriptsize} 
\rotate
\tablecaption{Statistics for Final Planetary Systems Formed from an Eccentric Initial Disk Mass Distribution}
\tablewidth{0pt} 
\tablehead{
\colhead{$\bf Run$} &
\colhead{$\it t$ (Myr)} &
\colhead{\bf $N_p$}   &
\colhead{\bf $N_m$}   &
\colhead{\bf $a_p$/$a^{\ast}_{c}$}   &
\colhead{$q_p$/$Q_{B}$}     & 
\colhead{$S_m$}     & 
\colhead{$S_s$}     & 
\colhead{\bf $S_d$} & 
\colhead{\bf $S_c$} & 
\colhead{\bf $S_r$} &
\colhead{\bf $m_l$ (\%)} &
\colhead{\bf $E/E_{0}$} & 
\colhead{$L/L_{0}$} & 
\colhead{\bf $L_Z/L_{Z_0}$}}
\startdata 

&&&&&&&&&&&&&&\\
CBecc\_.2\_.5\_.5\_a&200&3&0&1.89&4.22&0.481&33.0&0.0091&95.4&0.211&62.14&0.75&1.10&1.10\\
CBecc\_.2\_.5\_.5\_b&100&1&0&3.17&8.87&1.000&$\cdots$&0.0975&$\infty$&0.089&84.29&0.55&1.16&1.14\\
CBecc\_.2\_.5\_.5\_c&200&2&2&2.06&4.38&0.610&40.1&0.0048&115.1&0.213&62.50&0.66&1.18&1.17\\
CBecc\_.2\_.5\_.5\_d&200&2&0&1.76&3.82&0.652&48.2&0.0043&83.4&0.120&58.93&0.72&1.13&1.13\\
CBecc\_.2\_.5\_.5\_e&200&1&1&3.68&9.46&0.976&$\cdots$&0.0564&49.5&0.476&85.36&0.46&1.37&1.35\\
\bf{$\bar{x}$}&180 &1.8& 0.6&  2.51&   6.15& 0.744&  43.5&   0.0344& 91.7&    0.222&  70.64&  0.63&   1.19&   1.18\\

&&&&&&&&&&&&&&\\
CB\_.2\_.5\_.5\_a&146     &1&0&2.35&5.44&1.000&$\cdots$&0.0193&$\infty$&0.147&73.58&0.74&1.07&1.07\\
CB\_.2\_.5\_.5\_b &200     &2&0&2.25&4.65&0.890&57.2&0.0021&152.4&0.169&67.50&0.73&1.11&1.11  \\
CB\_.2\_.5\_.5\_c&154     &1&0&3.56&8.54&1.000&$\cdots$&0.0328&$\infty$&0.378&85.71&0.49&1.31&1.30\\
CB\_.2\_.5\_.5\_d &115     &1&0& 2.70&5.92&1.000&$\cdots$&0.0190&$\infty$&0.263&79.29&0.64&1.16&1.15\\
CB\_.2\_.5\_.5\_e &200  &2&0&1.78&5.20&0.590&32.4&0.0104&185.7&0.155&62.50&0.67&1.15&1.15\\ 
\bf{$\bar{x}$}&163&1.4  &   0.0&       2.53&   5.95  &  0.896  & 52.2&   0.0167& 179.0& 0.222&   73.72&  0.65&   1.16 &   1.16\\

&&&&&&&&&&&&&&\\
D& $\cdots$&0.20&0.20&0.40&0.60&0.40&0.60&0.40&0.80&0.40&0.40&0.20&0.20&0.20\\
Prob&$\cdots$&1.000&1.000&0.697&0.209&0.697&0.209&0.697&0.036&0.697&0.697&1.000&1.000&1.000\\

&&&&&&&&&&&&&&\\
\enddata 
\end{deluxetable}

\begin{figure}
\centering
\figurenum{1} 
\includegraphics[angle=90]{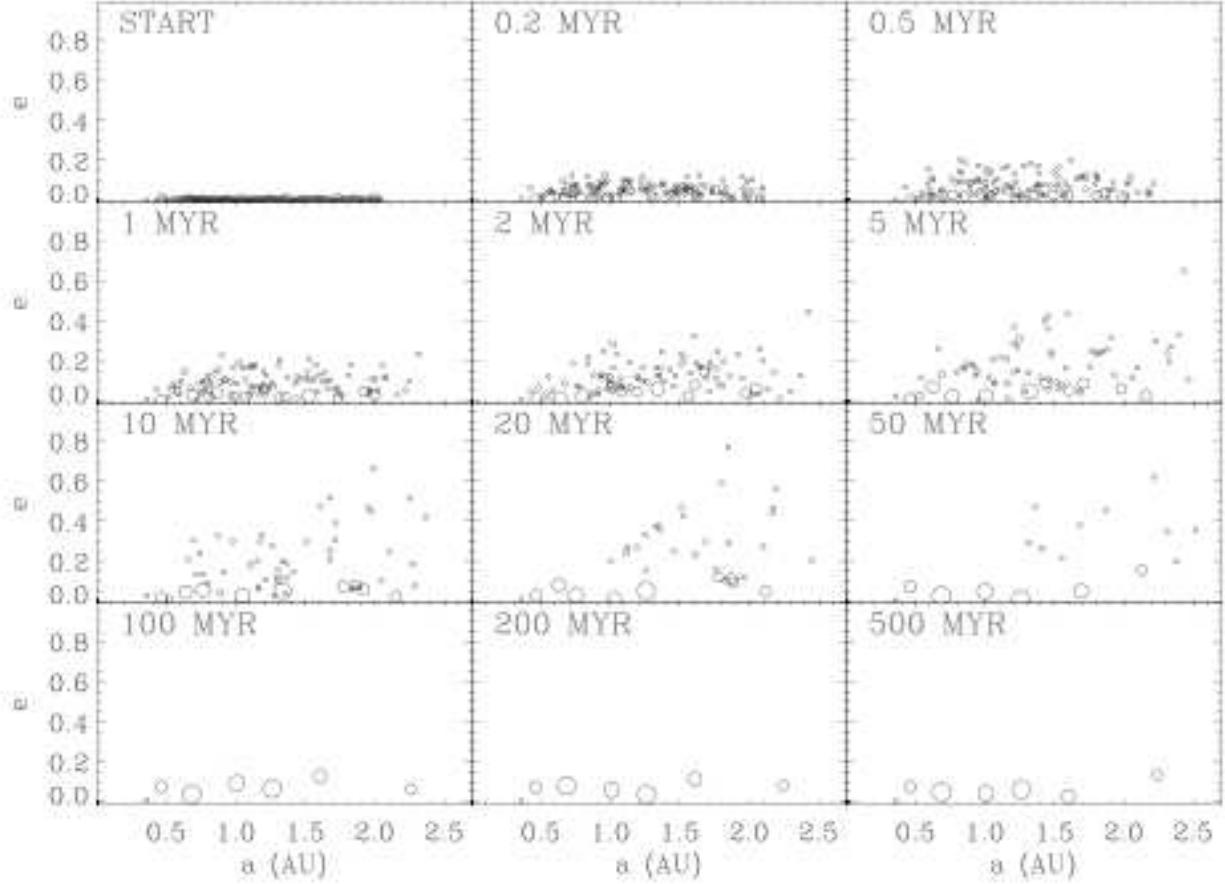}
\figcaption{The temporal evolution of the circumbinary disk in
  simulation CB\_.05\_0\_.5$^\dag$\_a.  For this simulation, the
  semimajor axis of the binary is $a_B$ = 0.05 AU, the binary orbit is
  circular ($e_B$ = 0) and each star has mass $M_{*}$ = 0.5
  M$_{\odot}$.  A single, Jupiter-like, giant planet is also included.
  The planetary embryos and planetesimals are represented by circles
  whose sizes are proportional to the physical sizes of the bodies.
  The locations of the circles show the orbital semimajor axes and
  eccentricities of the represented bodies relative to center of mass
  of the binary stars.  The initially dynamically cold disk heats up
  during the first 10 Myr, especially in the outer region, where the
  perturbations of the single giant planet included in this simulation
  are the greatest.  By 93 Myr into the simulation, six planets on low
  eccentricity orbits have formed, with one planetesimal remaining
  interior to their orbits.  All of these bodies survive for the
  remainder of the simulation.}
\end{figure}

\begin{figure}
\centering
\figurenum{2} 
\includegraphics[angle=90]{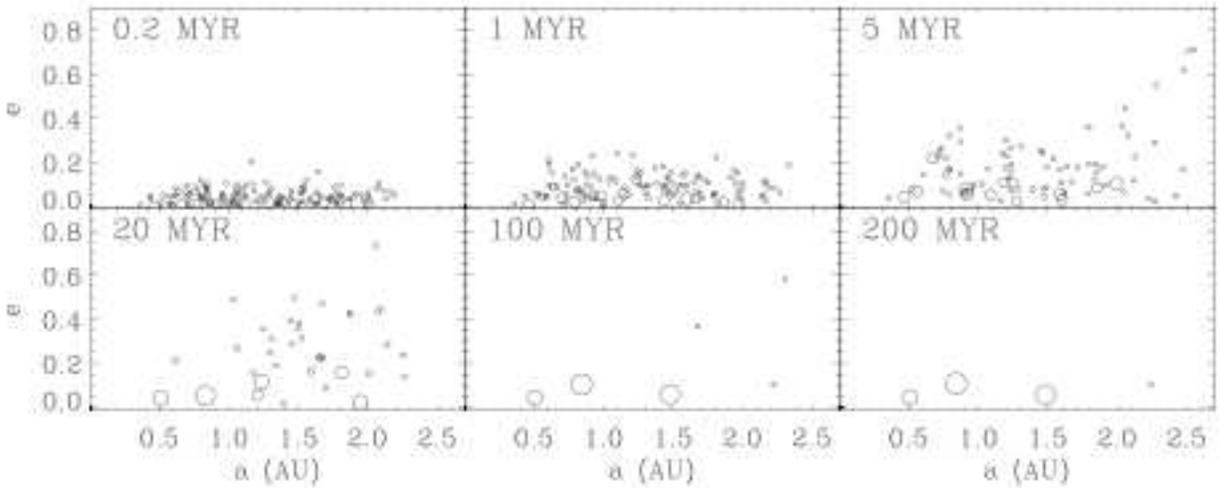}
\figcaption{The temporal evolution of simulation
  CB\_.05\_0\_.5$^\dag$\_d.  The masses and initial parameters of all
  bodies are identical to those used in the simulation displayed in
  Figure 1, aside from the shift of one planetesimal 3 meters forward 
  along its orbit.  Thus, the initial masses, semimajor axes and
  eccentricities are the same as those shown in the first panel of
  Figure 1.  The meanings of the symbols are as described in the
  caption to Figure 1.  Note that while this system seems
  qualitatively similar to the one represented in Figure 1 at early
  times, in this case only three terrestrial-mass planets survive.}
\end{figure}

\begin{figure}
\centering
\figurenum{3} 
\includegraphics[angle=90]{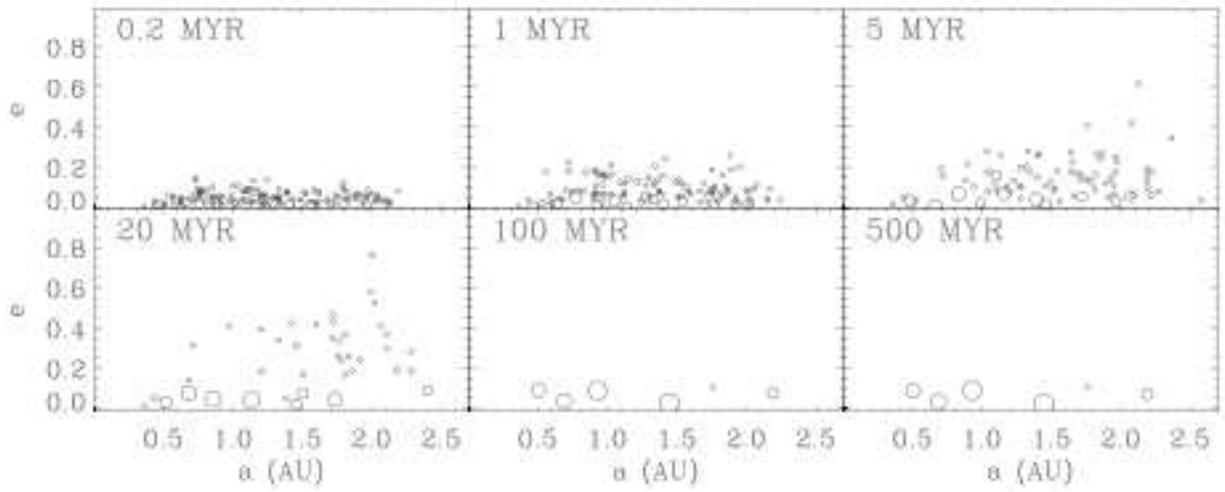}
\figcaption{The temporal evolution of simulation CB\_.05\_0\_.5\_c,
  which included both ``Jupiter'' and ``Saturn''.  Despite the
  addition of a second giant planet, the development of the system
  appears to be intermediate between the systems displayed in the
  previous two figures.  See the first panel in Figure 1 for the
  initial conditions and the caption of Figure 1 for explanation of
  the symbols.}
\end{figure}

\begin{figure}
\centering
\figurenum{4} 
\includegraphics[angle=90]{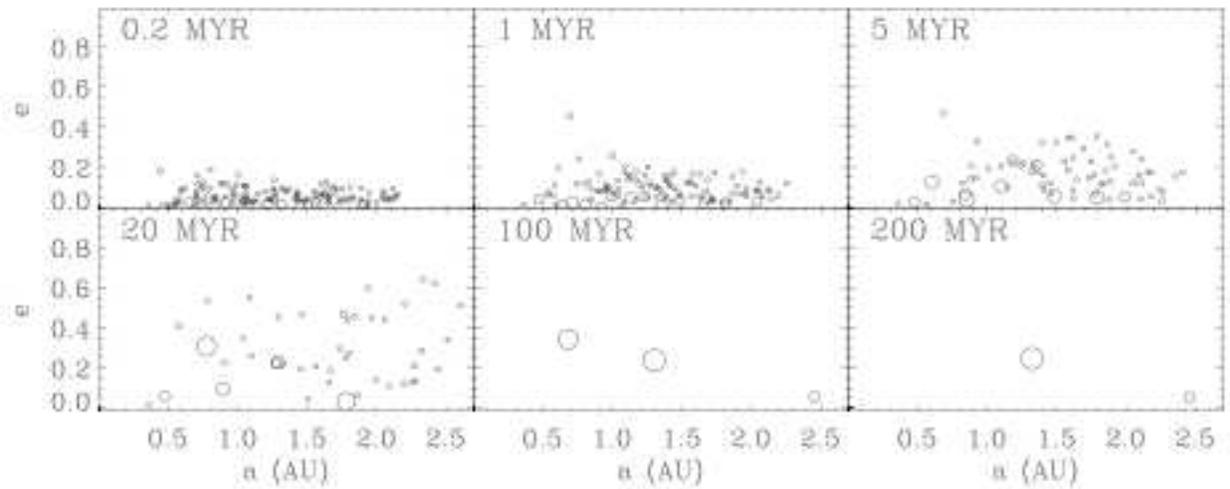}
\figcaption{The temporal evolution of simulation CB\_.075\_.33\_.5\_b.
  The larger binary semimajor axis ($a_B$ = 0.075 AU) and eccentricity
  ($e_B$ = 0.33) produce greater perturbations on the disk than those
  occurring in the systems shown in the previous figures.  The
  protoplanetary system is much more dynamically excited, especially
  in the inner regions, and a substantial amount of material is
  ejected, leaving only two planets.  See the first panel in Figure 1
  for the initial conditions and the caption of Figure 1 for
  explanation of the symbols.}
\end{figure}

\begin{figure}
\centering
\figurenum{5} 
\includegraphics[angle=90]{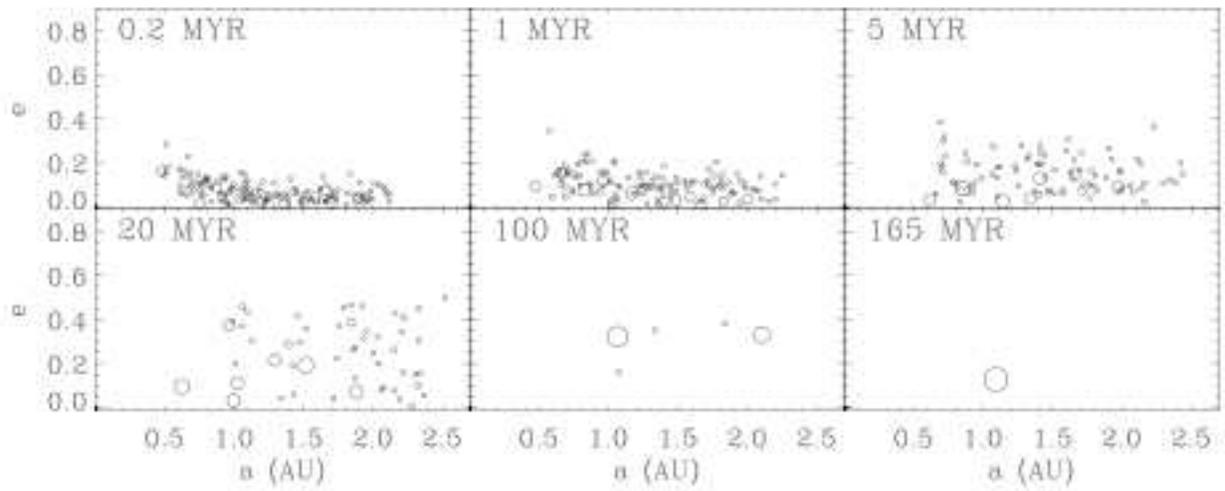}
\figcaption{The temporal evolution of simulation CB\_.1\_.8\_.2\_d.
  The effect of a larger stellar separation ($a_B$ = 0.1 AU) and
  higher binary eccentricity ($e_B$ = 0.8 AU), compared with previous
  figures, is apparent in the first panel where the inner part of the
  disk is already substantially excited by 200,000 yr.  Eccentricities
  remain high throughout the evolution, and by 164 Myr only one body
  remains in the terrestrial planet zone. See the first panel in
  Figure 1 for the initial conditions and the caption of Figure 1 for
  explanation of the symbols.}
\end{figure}
\clearpage

\begin{figure}
\centering
\figurenum{6} 
\includegraphics[angle=90]{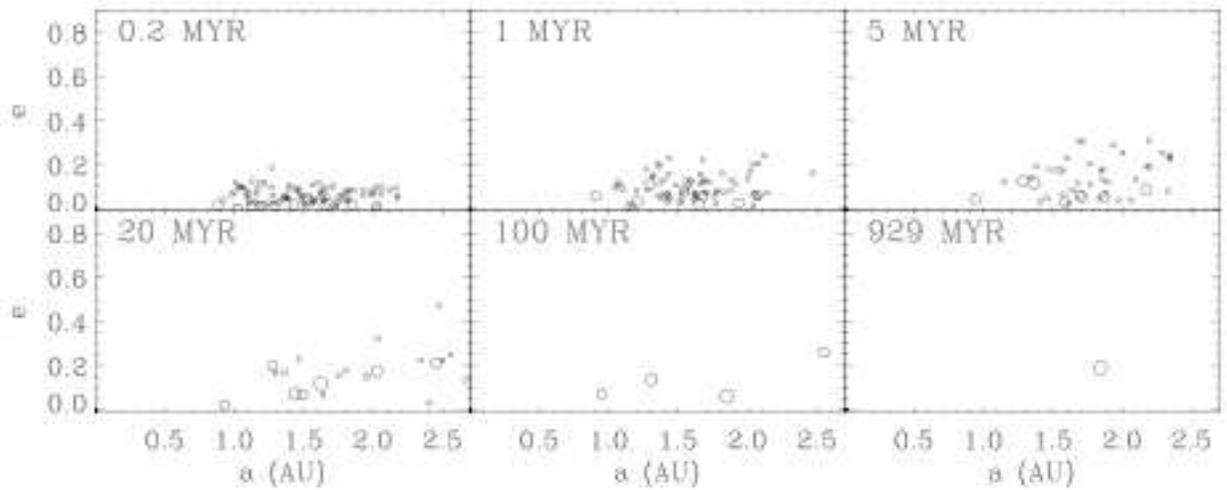}
\figcaption{The temporal evolution of simulation CB\_.4\_0\_.5\_c.
  With the binary on a circular orbit with semimajor axis equal to 0.4
  AU, the inner part of the disk is cleared out to beyond 0.8 AU by
  200,000 yr.  The system of four smallish planets that formed beyond
  0.9 AU by 100 Myr looks as if it might be stable.  However, the
  outermost planet was ejected at 110 Myr.  The remaining planets
  continued to interact until the orbits of the inner two planets
  crossed, resulting in the ejection of the innermost planet near 1 AU
  at 775 Myr.  The planet near 1.3 AU was ejected at 929 Myr, leaving
  just one body orbiting at 1.8 AU.  See the first panel in Figure 1
  for the initial conditions and the caption of Figure 1 for
  explanation of the symbols.}
\end{figure}

\begin{figure}
\centering
\figurenum{7} \epsscale{.85}
\plotone{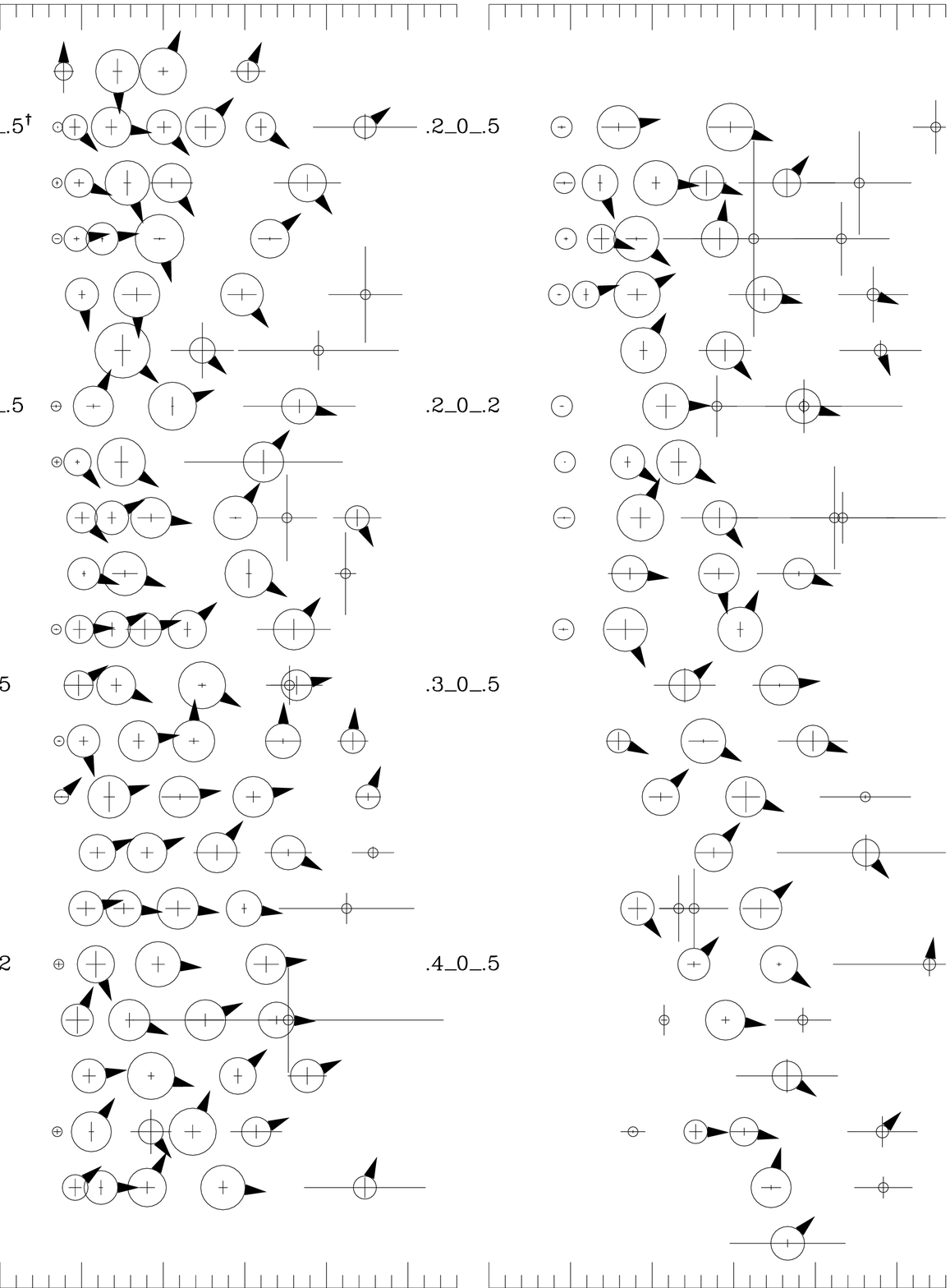}
\figcaption{}
\end{figure}

\begin{figure}
\centering
\figurenum{7} \epsscale{.75}
\figcaption{This figure displays the final planetary systems formed in
  all simulations which begin with binary stars on circular orbits
  ($e_B$ = 0) and with the initial disk coplanar with the stellar
  orbit.  Symbol sizes are proportional to planet sizes, the centers
  of the circles are positioned horizontally at the planet's semimajor
  axis, the horizontal bar represents the extent of the planet's
  radial excursions, and the vertical bar (which is proportional to
  the planet's inclination) shows its out of plane motion.  All
  orbital parameters are computed relative to the center of mass of
  the binary star system.  Arrows point in the direction representing
  the planet's rotational angular momentum; bodies that did not
  accrete do not have arrows attached because we have no information
  regarding their obliquity.  The terrestrial planets in our Solar
  System are shown in the upper left.  Modeled systems are grouped by
  common binary star parameters.  Note the diversity caused by chaos
  within individual groups, and also the general trends with increases
  in binary semimajor axis.  Also note that an additional planetesimal
  (which is not shown in this figure) remains in the third planetary
  system of set .2\_0\_.2 (simulations CB\_.2\_0\_.2\_c) at $a_p$ = 2.98
  AU, $e_p$ = 0.18 AU, and $i_p$ = 4.3$^{\circ}$ relative to the binary orbital plane.}
\end{figure}

\begin{figure}
\centering
\figurenum{8} \epsscale{.85}
\plotone{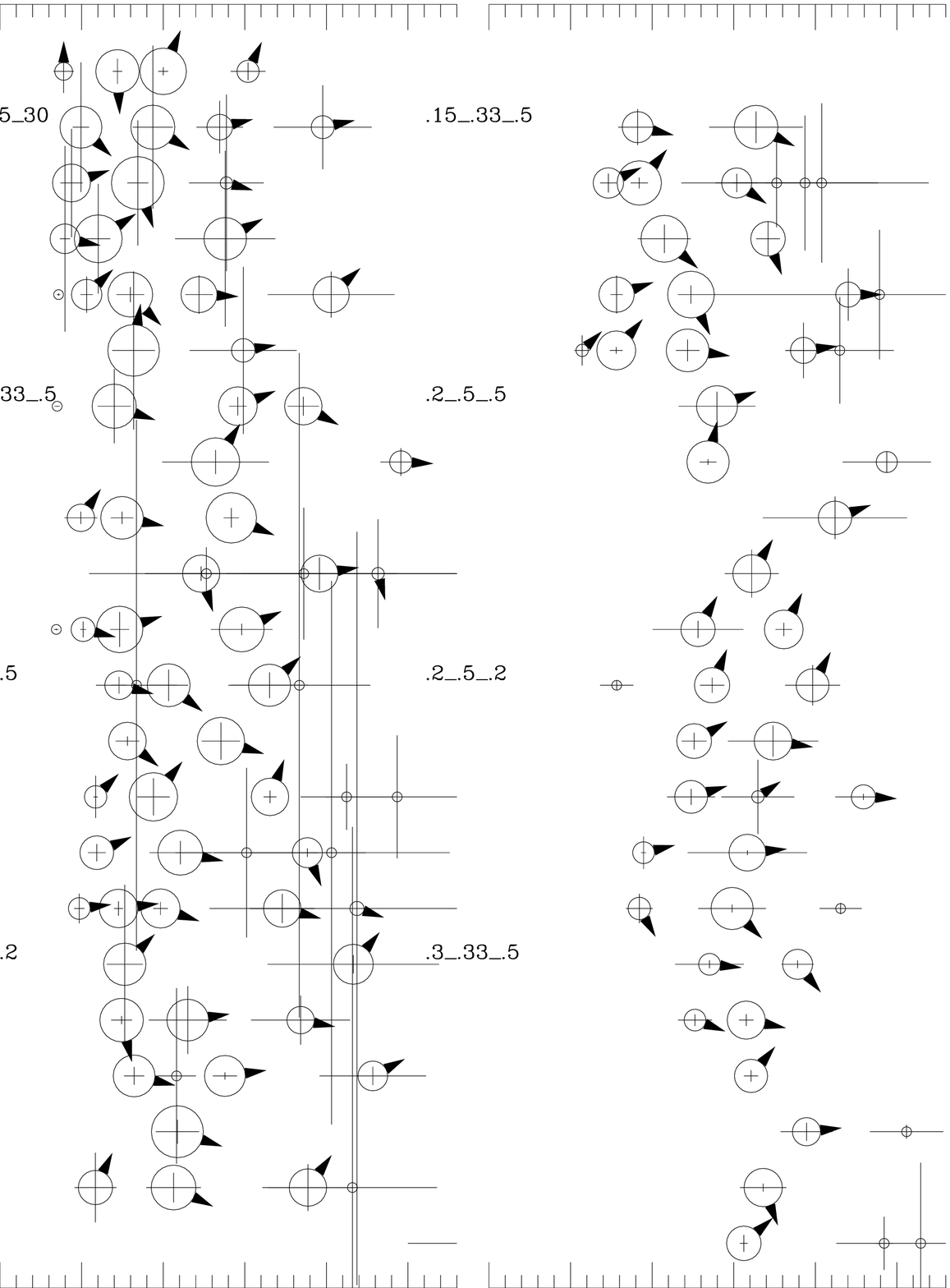}
\figcaption{}
\end{figure}

\begin{figure}
\centering
\figurenum{8} \epsscale{0.75} 
\figcaption{This figure displays the final planetary systems for the
  set of runs in which the disk was initially inclined by 30$^{\circ}$
  (labelled `.1\_0\_.5\_30'), and the simulations which begin with
  binary eccentricities $e_B >$ 0 (presented in order of increasing
  apastron $Q_B$).  Symbols are explained in the caption to Figure 7.
  Note that the planetary systems shown here generally are sparser and
  more diverse than those computed for binary stars on circular orbits
  (Fig. 7), but that again there is a wide range of outcomes due to
  the chaotic nature of planetary accretion dynamics.}
\end{figure}

\begin{figure}
\centering
\figurenum{9} \epsscale{0.6} 
\includegraphics[angle=90]{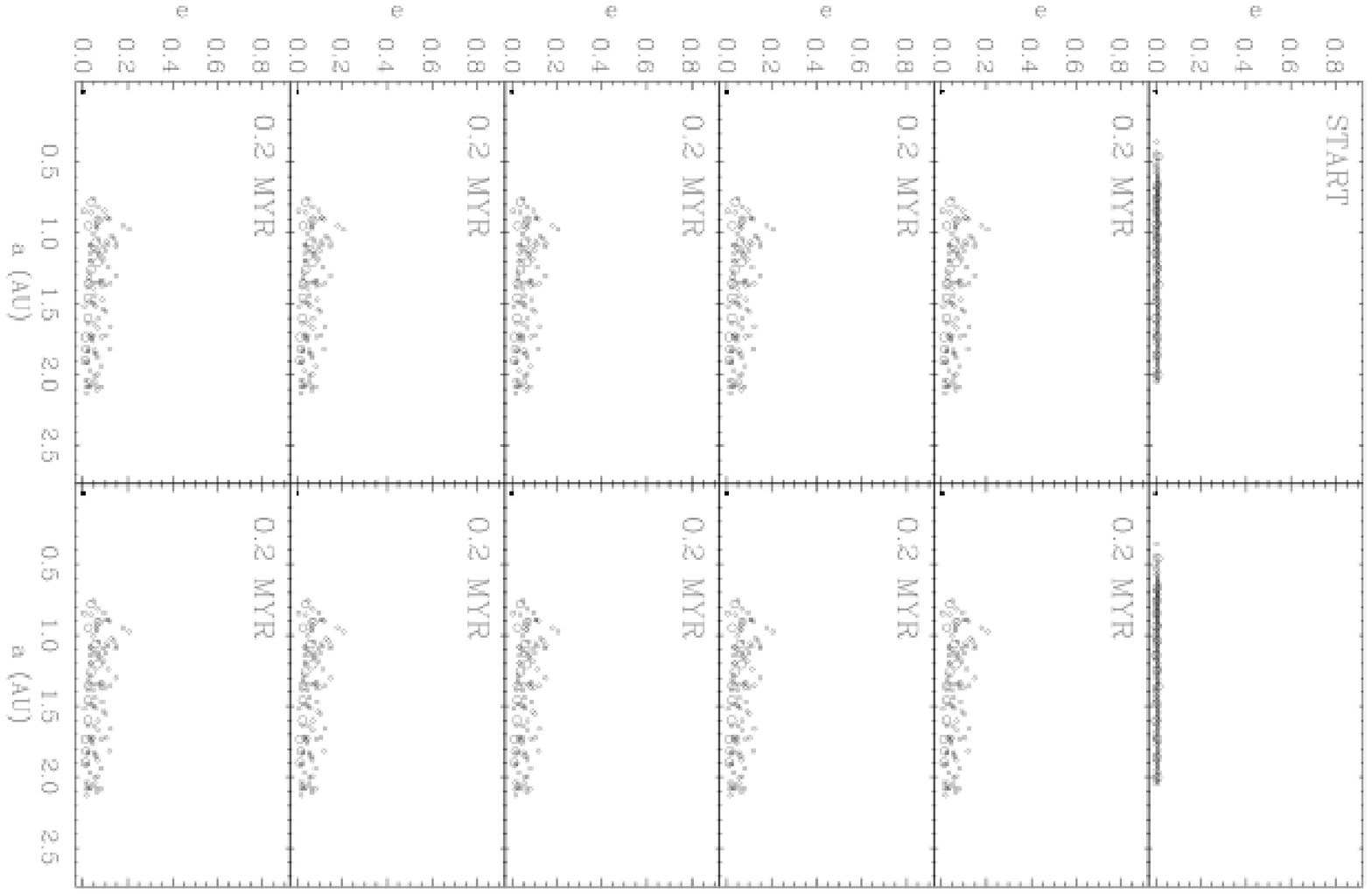}
\figcaption{This figure shows
  the early evolution of the circumbinary disk around binary stars
  with $a_B$ = 0.2 AU, $e_B$ = 0.5 and with equal masses of $M_{*}$ =
  0.5 M$_{\odot}$. Jupiter and Saturn are also included.  The left
  column shows our `nominal' disk (that which was used for almost all
  of the accretion simulations in this article), while the right
  column shows the disk that began with forced eccentricities.  The
  top row shows each circumbinary disk at the beginning of each
  simulation, and in subsequent rows the disk at $t$ = 200,000 years
  is shown for the five (a -- e) simulations.  The planetary embryos
  and planetesimals are represented by circles whose sizes are
  proportional to the physical sizes of the bodies as in Figs. 1 -- 6.
  The locations of the circles show the orbital semimajor axes and
  eccentricities of the represented bodies relative to center of mass
  of the binary stars.}
\end{figure}

\begin{figure}
\centering
\figurenum{10} \epsscale{0.4} 
 \plotone{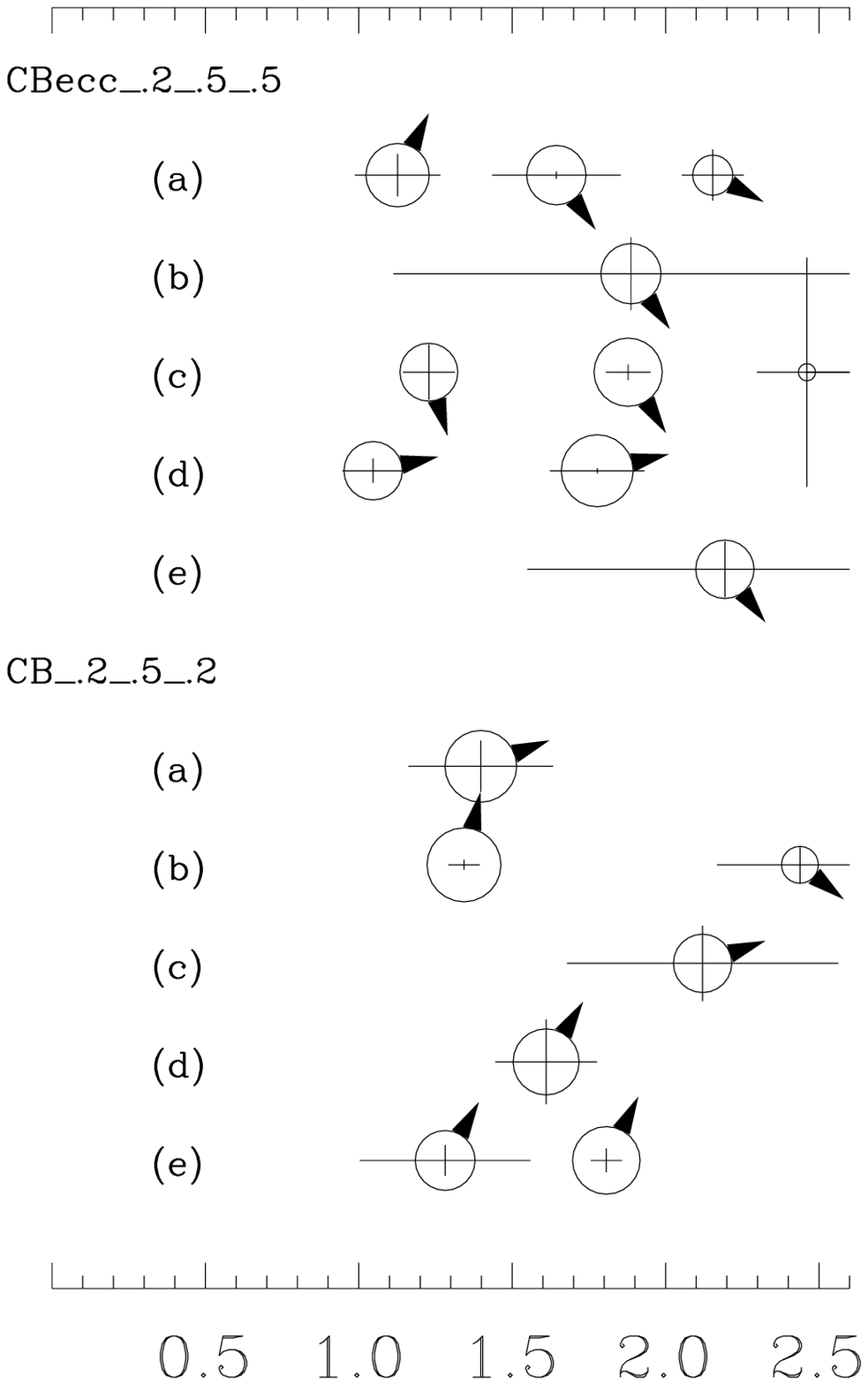}
\figcaption{The final
  planetary systems for the five simulations of set CBecc\_.2\_.5\_.5
  and the five runs of set CB\_.2\_.5\_.5.  Note that in run
  CBecc\_.2\_.5\_.5\_e, an additional planetesimal remains at $a_p$ =
  18.3 AU, with $e_p$ = 0.26 and $i_p$ = 28$^{\circ}$ relative to the
  stellar orbit.  As in
  Figures 7 and 8, symbol sizes are proportional to planet sizes, the
  centers of the circles are positioned horizontally at the planet's
  semimajor axis, the horizontal bar represents the extent of the
  planet's radial excursions, and the vertical bar (which is
  proportional to the planet's inclination) shows its out of plane
  motion.  All orbital parameters are computed relative to the center
  of mass of the binary star system.  Arrows point in the direction
  representing the planet's rotational angular momentum.  Although a
  net total of two more major planets were produced in the five CBecc
  runs, and two planetesimals also survived in these calculations,
  there is a large variation of planetary systems within each set as a
  result of chaos.  A statistical comparison between the two sets
  indicates that our choice of initial conditions for our `nominal'
  circumbinary disk is sufficient for the accretion simulations
  presented in this article.}
\end{figure}

\begin{figure}
\centering
\figurenum{11} \epsscale{0.5} 
\includegraphics[scale=.90,angle=90]{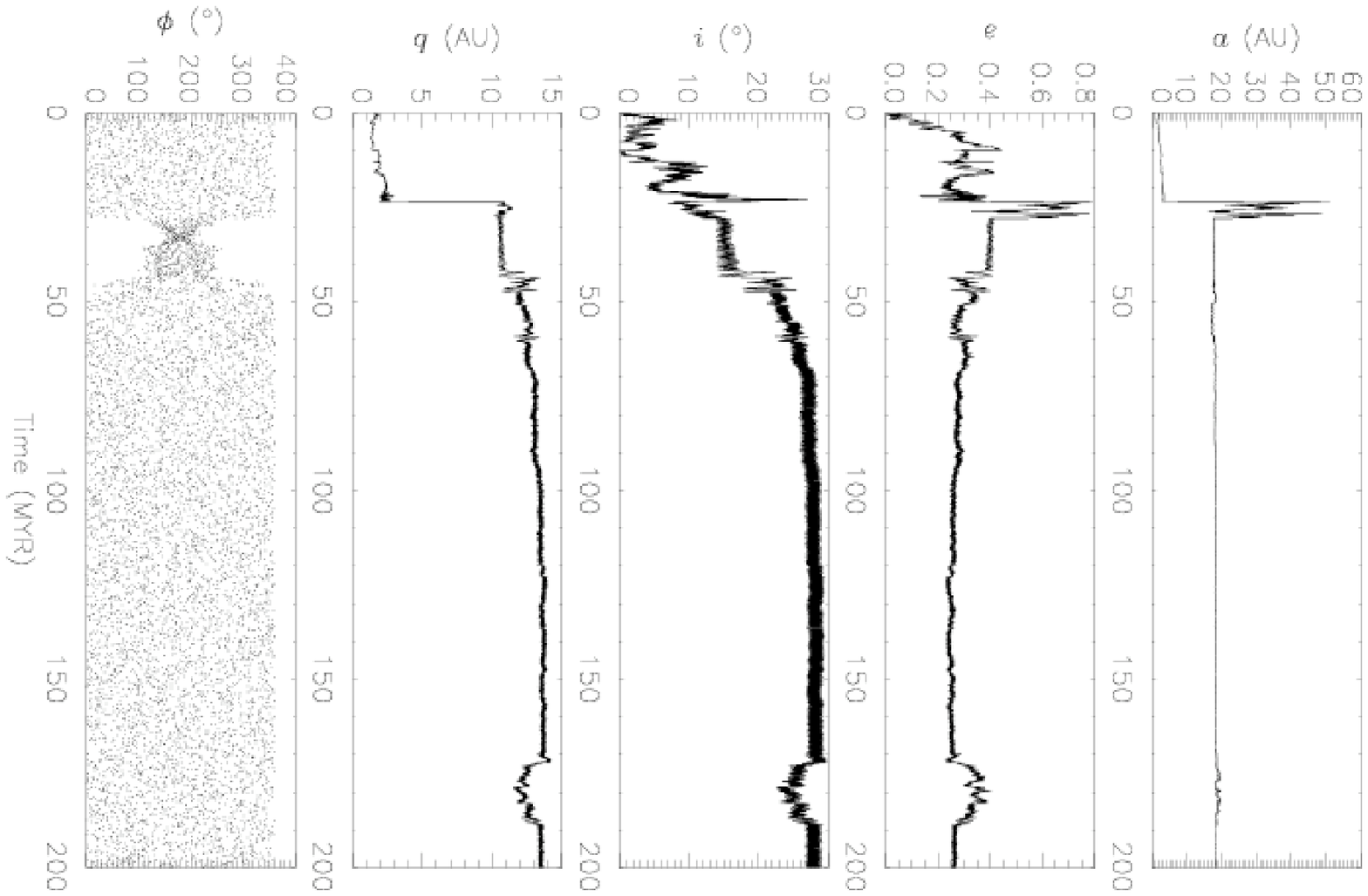}
\figcaption{Evolution of the
  planetesimal in simulation CBecc\_.2\_.5\_.5\_e that began with $a$
  = 1.83 AU, $e$ = 0.00015 and had $a$ = 18.3 AU, $e$ = 0.26 when the
  simulation was stopped at $t$ = 200 Myr.  (a) Semimajor axis.  (b)
  Eccentricity.  (c) Inclination.  (d) Periapse distance.  (e)
  Resonant angle for the 5:2 commensurability with the outer giant
  planet, $\phi$ $\equiv$ 5 $\lambda_{plmal}$ -- 2 $\lambda_{\saturn}$
  -- 3 $\tilde{\omega}_{plmal}$ where $\lambda$ = mean longitude and
  $\tilde{\omega}$ = longitude of periastron.  Note that once this
  planetesimal is scattered beyond the giant planets, its eccentricity
  and inclination are anticorrelated; this suggests that the Kozai
  mechanism plays an important role in the evolution of the
  planetesimal's periapse distance.}
\end{figure}

\end{document}